\documentclass[sigconf]{acmart}
\usepackage{algorithm,algpseudocode}
\usepackage{stmaryrd}
\usepackage{mdframed,multicol,multirow}
\usepackage{amsmath}
\usepackage{wasysym}
\usepackage{stfloats}
\usepackage{comment}
\usepackage{threeparttable}
\usepackage{url}
\usepackage{graphicx}
\usepackage{subfigure}
\usepackage{colortbl}
\usepackage[]{hyperref}
\usepackage{soul}
\usepackage{fancyhdr}
\usepackage{xcolor}
\usepackage{pifont}
\usepackage{nicematrix}
\usepackage{xspace}
\usepackage{enumitem}
\usepackage{marvosym}

\usepackage{pgfplots}

\usepackage{booktabs}
\usepackage{colortbl}
\usepackage{color}
\usepackage{tikz}
\usetikzlibrary{arrows}
\usetikzlibrary{math}
\usetikzlibrary{positioning}

\usetikzlibrary{quotes,angles}
\usetikzlibrary{calc}
\usetikzlibrary{decorations.pathreplacing}
\usetikzlibrary{shapes.arrows}
\usepackage{pgfplotstable}
\pgfplotsset{compat=1.7}
\usepackage{verbatim}

\definecolor{seagreen}{rgb}{0.18, 0.55, 0.34}
\definecolor{darkorange}{rgb}{1.0, 0.55, 0.0}
\definecolor{deeppink}{rgb}{1.0, 0.08, 0.58}
\definecolor{blue-violet}{rgb}{0.54, 0.17, 0.89}
\definecolor{brandeisblue}{rgb}{0.0, 0.44, 1.0}
\definecolor{carminered}{rgb}{1.0, 0.0, 0.22}

\def\eg{\textit{e.g.}}
\def\ie{\textit{i.e.}}
\def\etal{\textit{et~al. }}

\newcommand{\notesxf}[1]{{\color{purple}[\textsc{sxf:} #1]}}

\newcommand{\rewrite}[1]{{\color{green}{#1}}}
\newcommand{\notejbai}[1]{{\color{orange}[\textsc{jbai:} #1]}}

\newcommand{\mostree}{\textsf{Mostree}\xspace}

\newtheorem{theorem}{Theorem}
\newtheorem{lemma}{Lemma}
\newtheorem{definition}{Definition}

\usepackage{bm}

\newcommand{\myindent}{\vspace{0mm}\noindent}
\newcommand{\mysize}{\small}
\newcommand{\myspace}{\vspace{0mm}}
\newcommand{\myfigspace}{\vspace{0mm}}
\newcommand{\B}[1]{{\vec{{#1}}}\xspace}
\newcommand{\Twoboolshare}[1] {\llbracket #1 \rrbracket}
\newcommand{\Threeboolshare}[1]  {[ #1 ]^3}
\newcommand{\HMboolshare}[1] {\langle #1 \rangle}

\newcommand{\sh}[2] {\Twoboolshare{#1}_{#2}} 

\newcommand{\boolshare}[1] {\langle #1 \rangle}

\newcommand{\rnd}{\xleftarrow{\$}}

\newcommand{\rawtree}{{\B{\mathcal{T}}}}
\newcommand{\tree}{\B{\mathbf{T}}}
\newcommand{\treemac}{\B{\mathbf{M}}}
\newcommand{\feature}{\B{\mathbf{X}}}
\newcommand{\entree}{\mathcal{C}_{\mathcal{Y}}}
\newcommand{\enfeature}{\mathcal{C}_{\mathcal{X}}}

\newcommand{\idx}{\mathit{idx}}
\newcommand{\rdx}{\mathit{rdx}}

\newcommand{\f}[1]{\ensuremath{\mathcal{F}_{\mathsf{#1}}}}
\newcommand{\protocol}[1]{\ensuremath{{\rm \Pi}_{\mathsf{#1}}}}
\newcommand{\fsskey}[1]{\ensuremath{\mathsf{k}_{{#1}}^{\mathsf{dpf} }} }
\newcommand{\fsskeyy}[1]{\ensuremath{\mathsf{k}^{\mathsf{dpf}_{{#1}}^*}} }
\newcommand{\prfkey}[1]{\ensuremath{\mathsf{k}_{{#1}}^{\mathsf{prf}}}}
\newcommand{\mackey}{ \ensuremath{\alpha} }

\newcommand{\party}{{P}}

\newcommand{\MO}{\textsf{MO}\xspace}
\newcommand{\FO}{\textsf{FO}\xspace}
\newcommand{\CP}{\textsf{CP}\xspace}

\newcommand{\SIM}{\mathcal{S}}
\newcommand{\ADV}{\mathcal{A}}

\newcommand{\GG}{\mathbb{G}}

\newcommand{\ZZ}{\mathbb{Z}}

\newcommand{\FF}{\mathbb{F}}

\newcommand{\spdz}{SPDZ-tree\xspace}
\newcommand{\ucdt}{UCDT\xspace}

\newcommand{\symsec}{\ensuremath{\kappa}\xspace}
\newcommand{\statsec}{\ensuremath{\lambda}\xspace}

\renewcommand\footnotetextcopyrightpermission[1]{} 
\begin{document}

\title{\mostree : Malicious Secure Private Decision Tree Evaluation with Sublinear Communication}



\author{Jianli Bai}
\affiliation{%
  \institution{University of Auckland}
  \city{Auckland}
  \country{New Zealand}
}
\email{jbai795@aucklanduni.ac.nz}

\author{Xiangfu Song}
\authornote{Corresponding author}
\affiliation{%
  \institution{National University of Singapore}
  \country{Singapore}
}
\email{songxf@comp.nus.edu.sg}
\author{Xiaowu Zhang}
\affiliation{%
  \institution{CloudWalk Technology}
  \city{Beijing}
  \country{China}
}
\email{zhangxiaowu@cloudwalk.com}

\author{Qifan Wang}
\affiliation{%
  \institution{University of Auckland}
  \city{Auckland}
  \country{New Zealand}
}
\email{qwan301@aucklanduni.ac.nz}

\author{Shujie Cui}
\affiliation{%
  \institution{Monash University}
  \city{Melbourne}
  \country{Australia}
}
\email{shujie.cui@monash.edu}

\author{Ee-Chien Chang}
\affiliation{%
  \institution{National University of Singapore}
  \country{Singapore}
}
\email{changec@comp.nus.edu.sg}

\author{Giovanni Russello}
\affiliation{%
  \institution{University of Auckland}
  \city{Auckland}
  \country{New Zealand}
}
\email{g.russello@auckland.ac.nz}

\begin{abstract}

A private decision tree evaluation~(PDTE) protocol allows a feature vector owner~(\FO) to classify its data using a tree model from a model owner~(\MO) and only reveals an inference result to the \FO. This paper proposes \mostree, a PDTE protocol secure in the presence of malicious parties with sublinear communication. We design \mostree in the three-party honest-majority setting, where an (untrusted) computing party~(\CP) assists the \FO and \MO in the secure computation. We propose two low-communication oblivious selection~(OS) protocols by exploiting nice properties of three-party replicated secret sharing~(RSS) and distributed point function. \mostree combines OS protocols with a tree encoding method and three-party secure computation to achieve sublinear communication. We observe that most of the protocol components already maintain \textit{privacy} even in the presence of a malicious adversary, and what remains to achieve is \textit{correctness}. To ensure correctness, we propose a set of lightweight consistency checks and seamlessly integrate them into \mostree. As a result, \mostree achieves sublinear communication and malicious security simultaneously. We implement \mostree and compare it with the state-of-the-art. Experimental results demonstrate that \mostree is efficient and comparable to semi-honest PDTE schemes with sublinear communication. For instance, when evaluated on the MNIST dataset in a LAN setting, \mostree achieves an evaluation using approximately 768 ms with communication of around 168 KB.
\end{abstract}



\maketitle






\section{Introduction}\label{sec:introduction}

Decision trees have found extensive use in various real-world applications, including spam filtering~\cite{bratko2006spam}, credit risk assessment~\cite{koh2006two}, and disease diagnosis~\cite{podgorelec2002decision}.
In many scenarios, the tree model owner and feature owner are distinct parties, and neither of them wants to disclose data to the other due to commercial or privacy concerns.

Private decision tree evaluation~(PDTE) protocols allow a feature owner (\FO) to learn a classification result evaluated using a decision tree (DT) from a model owner (\MO) without revealing anything more.  
Recently, many PDTE protocols \cite{bost2015machine,kiss2019sok,brickell2007privacy,joye2018private,wu2016privately,tai2017privacy,ma2021let,damgaard2019new} have been proposed with different security and efficiency trade-offs.
Ideally, a PDTE protocol should achieve sublinear communication in the tree size. 
However, as shown in Table~\ref{table::comparison}, most existing PDTE protocols~\cite{bost2015machine,kiss2019sok,brickell2007privacy,joye2018private,wu2016privately,tai2017privacy} require linear communication, which can be impractical for real-world applications since commercial DTs typically contain thousands or millions of nodes~\cite{kotsiantis2013decision}. 
Moreover, almost all existing PDTE protocols are only secure against semi-honest adversaries, where the adversary honestly follows the protocol specification. 
Indeed, the adversary could maliciously behave.
Achieving private and correct tree evaluation in the presence of a malicious adversary is vital since PDTE protocols are usually used for high-sensitive applications, \eg, healthcare or financial services. 
Only the works proposed in \cite{wu2016privately, tai2017privacy,ma2021let,damgaard2019new} achieve security against malicious adversaries. In particular, the works~\cite{wu2016privately,tai2017privacy,ma2021let} only achieve security against a malicious \FO, which we call one-side malicious security. 
Besides, all four works require linear communication costs, limiting their scalability. 

In this paper, we propose \mostree, a PDTE protocol that simultaneously achieves security against malicious adversaries and sublinear communication. 
\mostree considers a three-party honest-majority setting, where a malicious adversary can compromise one party, and the compromised party could be anyone in the system. 
The main application for \mostree can be found in cloud-assisted privacy-preserving machine learning~(PPML) service, as considered by Sharemind ({\url{https://sharemind.cyber.ee}), TFEncrypted ({\url{https://tf-encrypted.io}) and SecretFlow ({\url{https://www.secretflow.org.cn/docs/secretflow/latest/en-US}).
\mostree protects both the tree model and the queried features from all parties using replicated secret sharing (RSS). 
During the tree evaluation, protecting which node is being accessed in each level, \ie~ the tree access pattern, is also imperative since the classification result is highly relevant to the tree path. In particular, if a model holder knows which path is accessed for each query, it can learn the classification result directly, which should be forbidden to protect the feature owner's privacy. 
\mostree uses oblivious selection (OS) protocols to hide the access pattern from all parties, which allows the three parties to traverse the decision tree collaboratively and obliviously, without learning which node is being touched. 
We design two OS protocols. 
The first OS protocol is purely based on RSS sharings, achieving constant online communication and linear offline communication. 
We then propose the second OS protocol by applying distributed point function (DPF)~\cite{ishai2007evaluating} over RSS sharings, achieving constant online and sublinear offline communication.

\myspace
\begin{table*}[!htp]
    \caption{Summary of Some Existing PDTE Protocols}
	\vspace{-2mm}
	\label{table::comparison}
	\centering 
	\begin{threeparttable}
    \begin{tabular}{cccccccc}
		\toprule
        \textbf{Protocol} & \textbf{Comparison} & \textbf{Communication} &\textbf{Sublinear} & \textbf{Leakage} & \textbf{Security} & \textbf{Corruption} & \textbf{Parties} \\
        
		\hline
		Bost \etal~\cite{bost2015machine} & $\lceil m/2 \rceil$ & $O(n + m)$ & $\times$ & $m$  & \Circle & 1-out-of-2 & 2PC\\
		~ Kiss \etal~\cite{kiss2019sok}(GGG) & $d$ & ${O}(\overline{m}\ell)$ & $\times$ & $\overline{m},d$  & \Circle & 1-out-of-2 & 2PC \\
		~ Kiss \etal~\cite{kiss2019sok}(HHH) & $\lceil m/2 \rceil$ & ${O}((n + m)\ell)$ & $\times$ & $m$  & \Circle & 1-out-of-2 & 2PC \\
		~ Brickell \etal~\cite{brickell2007privacy} & $d$ & ${O}((n + m)\ell)$ & $\times$ & $m$  & \Circle & 1-out-of-2 & 2PC \\
        ~ Joye \etal~\cite{joye2018private} & $d$ & $O(d(\ell+n)+2^d)$ & $\times$ & $d$  & \Circle & 1-out-of-2 & 2PC \\
		~ Tueno \etal~\cite{tueno2019private}(ORAM)& $d$ & $O(d^4\ell)$ & $\surd$ & $d$  & \Circle & 1-out-of-2 & 2PC \\
		~ Bai \etal~\cite{bai2022scalable} & $d$ & $O(dn\ell)$ & $\surd$ & $m,d$  & \Circle & 1-out-of-2 & 2PC \\
        ~ Wu \etal~\cite{wu2016privately} & $2^d$ & $O(2^{d}+ (n + m)\ell)$ & $\times$ & $m,d$   & \RIGHTcircle & 1-out-of-2 & 2PC \\
		~ Tai \etal~\cite{tai2017privacy} & $\lceil m/2 \rceil$ & ${O}((n + m)\ell)$ & $\times$ & $m$  & \RIGHTcircle & 1-out-of-2 & 2PC\\
        ~ Ma \etal~\cite{ma2021let} & $d$ & $O(dn\ell)$ & $\sqrt{}\mkern-9mu{\smallsetminus}$ & $m,d$   & \RIGHTcircle & 1-out-of-2 & 2PC\\
		\hline \hline
		~Ji \etal~\cite{ji2021uc} & $d$ & $O(d(\log n+\log m+\ell))$ & $\surd$ & $m,d$  & \Circle & 1-out-of-3 & 3PC \\
		~Damg{\aa}rd \etal~\cite{damgaard2019new} & $m$ & $O(2^dn\ell)$ & $\times$ & $\overline{m},d$  & \CIRCLE & 2-out-of-3 & 3PC\\
        ~\textbf{\mostree} & $d$ & $O(dn\ell\log m)$  & $\surd$  & $m,d$ & \CIRCLE & 1-out-of-3 & 3PC\\
        \bottomrule
	\end{tabular}

	\itshape \rm {\textit{Comparison} denotes the number of secure comparisons needed;
    \textit{Parameters}: $m$: number of tree nodes, $\overline{m}$: number of tree nodes in a depth-padded tree, see~\cite{kiss2019sok}, $n$: dimension of a feature vector, $d$: the longest depth of a tree, $\ell$: bit size of feature values. \textit{Symbols}: $\times$: no, $\surd$: yes, $\sqrt{}\mkern-9mu{\smallsetminus}$: partially support: linear offline communication, sublinear online communnication; 
    \Circle: semi-honest, \RIGHTcircle: one-side malicious, \CIRCLE: malicious. 
	}
    \end{threeparttable}
\end{table*}

The remaining challenge is how to achieve security against malicious adversaries while ensuring sublinear communication. 
We exploit the fact that most of our proposed semi-honest protocol components already maintain \textit{privacy} in the presence of a malicious adversary; what we need to ensure is \textit{correctness}.\footnote{Informally, privacy requires that a protocol reveals nothing except the protocol output and any allowed information. Correctness requires the computation to be done correctly.}
To this end, we propose a set of lightweight consistency check techniques.
Notably, all of our check mechanisms are efficiency-oriented by exploiting nice properties of underlying primitives, and they bootstrap existing RSS-based security mechanisms (\eg, low-level RSS-based ideal functionalities) to ensure correctness and introduce low overhead. 
By integrating these checks, \mostree simultaneously achieves sublinear communication and security against malicious adversaries.

We implement \mostree over different datasets and report efficiency under different network settings. Our results demonstrate that \mostree is highly competitive to the PDTE protocol proposed in~\cite{ji2021uc}, which is the latest and most efficient existing solution designed in semi-honest settings. 
In MNIST testing within a LAN environment, \mostree requires only $4\times$ online communication and $4\times$ online computation compared to~\cite{ji2021uc}.
Compared with the malicious security work given in~\cite{damgaard2019new}, 
\mostree reduces up to around 311$\times$ and about 4$\times$ in communication and computation, respectively. 
Furthermore, our experiments illustrate the scalability of \mostree, particularly for large trees with high dimensions, due to its sublinear communication property. 


\noindent\textbf{Contributions}. We summarize our contributions as follows: 
\begin{itemize}[leftmargin=10pt]
    \item[$\bullet$] We propose two oblivious selection protocols over a three-party setting. The first protocol is based purely on RSS, and the second is on DPF and RSS.
    Both of them are with low overhead, \eg, sublinear communication. 
    The proposed OS protocol may apply to applications in other areas, \eg, secure database processing.  
    
    \item[$\bullet$] 
    We enhance OS protocols with malicious security by proposing lightweight consistency check techniques.
    We carefully combine the proposed OS protocol, efficient consistency check, and three-party secure computation to design \mostree. \mostree achieves sublinear communication and malicious security in the three-party honest-majority setting.
    To our best knowledge, \mostree is the first PDTE protocol that simultaneously achieves the above two properties.
    
    \item[$\bullet$] We implement \mostree and measure its performance. The experiment results show \mostree is highly communication-efficient compared with the state-of-the-art.  
\end{itemize}

\section{Related Work} 
\label{sec:relatedwork}
We categorize PDTE protocols based on the adversary models: semi-honest PDTE protocols and malicious PDTE protocols. Table~\ref{table::comparison} provides a comprehensive comparison of some representative PDTE schemes, \eg, PDTE in two-party settings \cite{bost2015machine, kiss2019sok,brickell2007privacy,joye2018private,tueno2019private,bai2022scalable,wu2016privately,tai2017privacy,ma2021let} and PDTE in three-party settings \cite{ji2021uc, damgaard2019new}, taking into account their performance and security guarantees. 



\myindent \textbf{Semi-honest PDTE Protocols.} Most existing PDTE protocols, \eg,~ \cite{ishai2003extending,brickell2007privacy,barni2009secure,bost2015machine,wu2016privately,tai2017privacy,de2017efficient,kiss2019sok}, are only secure against semi-honest adversaries. 
Moreover, they come with heavy computation and/or communication overhead.
In~\cite{ishai2007evaluating}, Ishai and Paskin evaluate DTs via homomorphic encryption~(HE), which brings huge computation burden and linear communication cost. 
Protocols proposed in~\cite{brickell2007privacy} and~\cite{barni2009secure} require linear communication overhead since the tree transferred can never be re-used due to the access pattern leakage. 
Later, Bost~\etal~\cite{bost2015machine} encode the tree into a high-degree polynomial, simplifying the communication process but requiring costly fully HE. 
Wu~\etal~\cite{wu2016privately} avoid computing expensive polynomials used in~\cite{bost2015machine} by sending an encrypted permuted tree to \FO. Tai \etal~\cite{tai2017privacy} adopt the same strategies as~\cite{wu2016privately}, but rather than transmitting the entire tree, \MO computes and sends a value for each path, which they call path cost. By doing so, they save communication costs. 
Following path costs concept from~\cite{tai2017privacy}, many subsequent works~\cite{de2017efficient}~\cite{kiss2019sok} have been proposed. The work proposed by Kiss \etal ~\cite{kiss2019sok} mainly concentrates on exploring the influence of different combinations of HE and MPC on the performance of PDTE.  

\myindent \textit{Sublinear complexity.}
As indicated in Table~\ref{table::comparison}, most PDTE protocols have linear complexity in both communication and computation. This limitation renders them impractical for evaluating large decision trees containing millions of nodes \cite{catlett1991overprvning}.
Researchers are keen to explore PDTE protocols with overhead sublinear to the tree size. 
Tueno~\etal~\cite{tueno2019private} design the first sublinear protocol with semi-honest security in the two-party setting. Their idea is to represent the tree in an array. The tree construction allows \MO to obliviously select the tree node in each tree level if the node index is shared between \FO and \MO. The selection process is called Oblivious Array Index (OAI). They instantiate the OAI approach by ORAM, which results in $O(d^4)$ communication cost and $d^2$ rounds for complete trees. Joye and Salehi~\cite{joye2018private} also work on a semi-honest two-party scenario but employ a different strategy to achieve oblivious selection. They observe there is only one node to be selected from each tree level, and thus their method is obliviously selecting a node from $n=2^l$ nodes where $l$ is the sitting tree level. For the whole tree evaluation, only $d$ comparisons are required, where $d$ is the depth of the tree. However, the communication is still linear to the tree size.
Ma \etal \cite{ma2021let} follows the idea of~\cite{brickell2007privacy} to send the encrypted tree to \FO. Rather than using a complex garbled circuit, they employ secret sharing to protect the tree. In each level, \FO performs Oblivious Transfer (OT) with \MO to retrieve the nodes to be evaluated, which optimizes both communication and computation overhead of~\cite{brickell2007privacy}. However, similar to~\cite{brickell2007privacy}, \FO can still learn extra information from the access pattern leakage. The work proposed in~\cite{ji2021uc} achieves sublinear communication by leveraging function secret sharing. However, this work is only designed to defend against semi-honest adversaries.  

\myindent \textbf{Malicious Secure PDTE Protocols.}
Existing works \cite{wu2016privately,tai2017privacy,ma2021let} can protect the model from a malicious \FO. The idea from~\cite{wu2016privately,tai2017privacy} is \FO additionally sends zero-knowledge proofs (ZKP) to \MO to prove the correctness of inputs. Both protocols result in heavy linear computation and communication overhead. 
In work~\cite{ma2021let}, Ma \etal replace Garbled Circuit (GC) protocol with its maliciously secure version~\cite{asharov2013more} and employ commitment and ZKP to constrain the parties' behavior. However, this protocol still suffers from linear communication costs because the tree can never be reused after evaluation. All the above three works cannot guarantee correctness when \MO is malicious.
Damg{\aa}rd \etal~\cite{damgaard2019new} present a PDTE protocol using SPD$\mathbb{Z}_{2^k}$ in the dishonest majority setting (up to $n-1$ corruption out of $n$ parties). Similar to previous works~\cite{wu2016privately,tai2017privacy,de2017efficient}, they have \MO and \FO to securely perform attribute selection and comparison for each tree node, resulting in $O(mn)$ computation and communication,  where $m$ is the tree size, and $n$ is the feature size.
Notably, their protocol can protect both privacy and correctness and achieves higher security than ours since they work in the dishonest majority setting.  

\myindent\textbf{Recent DPF-over-RSS Techniques}. 
We note several constructions that utilize DPFs over RSS~\cite{dauterman2022waldo,ji2021uc,wagh2022pika,vadapalli2022duoram} were proposed recently. 
Waldo~\cite{dauterman2022waldo} use DPFs over RSSs to design a privacy-preserving database query. However, Waldo relies on honest clients to generate and distribute the DPF keys, whereas, in \mostree, the keys are generated and distributed by a possibly malicious party, requiring additional checks to ensure correctness. 
The DPF-over-RSS technique is also used in \cite{ji2021uc} and \cite{vadapalli2022duoram}, but they only achieve semi-honest security. 
Pika~\cite{wagh2022pika} uses DPFs over ring-based RSS and achieves malicious security.
Our scheme operates specifically on boolean-based RSS, which enables the design of more efficient error detection mechanisms. We will show more in the following sections.

\section{Background}\label{sec:background}

\myspace
\subsection{Decision Trees Evaluation}
A decision tree is usually represented as a binary tree where its inner nodes are \textit{decision nodes} and its leaves are \textit{classification nodes}. 
A decision node consists of a threshold and the index of the corresponding feature attribute. A classification node contains a classification label. 
Given a feature vector, DTE starts from the root.
It compares a feature value with the threshold value and decides which child to visit based on the result (\ie, 1 for the left child and 0 for the other).
The evaluation continues until it reaches a leaf, from which the evaluation outputs a label as the classification result.

\subsection{Cryptographic Primitives} \label{Pre:3pc-rss}
\myindent\textbf{Notations}. 
We use $P_i$ to denote the $i$th party, where $i \in \{0,1,2\}$ and we write $P_{i-1}$ and $P_{i+1}$ as its ``previous'' and ``subsequent'' parties, respectively. Typically, $P_{i-1}$ is $P_2$ when $i=0$ and $ P_{i+1}$ is $P_0$ when $i=2$.
We interchangeably use $\mathbb{F}_2^k$ and $\mathbb{F}_{2^k}$ to represent the data in $\{0,1\}^k$, depending on the context. 
Addition in $\mathbb{F}_2^k$ and $\mathbb{F}_{2^k}$ corresponds to bit-wise XOR operation. 
We write $\mathbb{F}_{2^k} \cong \mathbb{F}[X]/f(X)$ for some monic, irreducible polynomial $f(X)$ of degree $k$.
We denote the set $\{0,\cdots, j-1\}$ as $[j]$.
Given two vectors $\B{x}$ and $\B{y}$, we use $\B{x} \odot \B{y}$ to denote inner-product computation between $\B{x}$ and $\B{y}$.

\myindent\textbf{Secret Sharing}.
We use secret sharing for secure computation.

\begin{itemize}[leftmargin=10pt]
    \item[$\bullet$] \textbf{$(^n_n)$-sharing $\Twoboolshare{x}$.} 
    We use $\Twoboolshare{x}$ to denote $x \in \FF$ is shared in $n$ parties by $(^n_n)$-sharing, where $\party_{i}$ holds a share $\sh{x}{i} \in \FF$ satisfying $x = \sum_{i \in [n]} \sh{x}{i}$. We use $n=2$ and 3 in this paper. 
    
    \item[$\bullet$] \textbf{$(^3_2)$-sharing $\HMboolshare{x}$.}
    We use $\HMboolshare{x}$ to denote $x$ is shared by $(^3_2)$-sharing, 
    also known as \textit{replicated secret-sharing}~(RSS). 
    In RSS sharing, we denote $\HMboolshare{x} = (x_0, x_1, x_2)$, where each party $\party_i$~($i \in \{0,1,2\}$) holds two shares $(\sh{x}{i}, \sh{x}{i-1})$ such that $\sh{x}{0}+\sh{x}{1}+\sh{x}{2} = x$. 
    Naturally, given a public value $v$, it can be shared as $\HMboolshare{v} = (0,0,v)$.
    Different from $(^3_3)$-sharing, any two parties in $(^3_2)$-sharing can reconstruct the secret.
\end{itemize}

We extend the above definition to vectors. 
We use $\B{x} \in \FF^m$ to denote an $m$-dimensional vector.
Accordingly, we use $\Twoboolshare{\B{x}}$ and $\HMboolshare{\B{x}}$ to denote a $(^n_n)$-sharing and $(^3_2)$-sharing of a vector $\B{x}$, respectively.

\myindent\textbf{Secure Computation over RSS Sharing}. 
RSS sharing supports the following (semi-honest) addition and multiplication operations: 
\begin{itemize}[leftmargin=10pt]
    \item[$\bullet$] $\HMboolshare{z} \leftarrow \HMboolshare{x} + \HMboolshare{y}$: For $i \in [3]$, $\party_i$ computes $(\sh{z}{i} = \sh{x}{i}+\sh{y}{i}, \sh{z}{i-1} = \sh{x}{i-1}+\sh{y}{i-1})$. 
    
    \item[$\bullet$] $\HMboolshare{z} \leftarrow \HMboolshare{x} + c$: 
    $\party_0$ computes $(\sh{z}{0}, \sh{z}{2}) = (\sh{x}{0}+c, \sh{x}{2})$; $\party_1$ computes $(\sh{z}{1}, \sh{z}{0}) = (\sh{x}{1}, \sh{x}{0}+c)$; and $\party_2$ computes $(\sh{z}{2}, \sh{z}{1}) = (\sh{x}{2}, \sh{x}{1})$. 
    
    \item[$\bullet$] $\HMboolshare{z} \leftarrow c\cdot \HMboolshare{x}$: 
    For $i \in [3]$, $\party_i$ computes $(\sh{z}{i}, \sh{z}{i-1}) = (c \cdot \sh{x}{i}, c  \cdot\sh{x}{i-1})$.  
    
    \item[$\bullet$] $\HMboolshare{z} \leftarrow \HMboolshare{x} \cdot \HMboolshare{y}$:
    $\party_i$ computes $\sh{t}{i} \leftarrow \sh{x}{i}\cdot \sh{y}{i} + \sh{x}{i-1} \cdot \sh{y}{i} + \sh{x}{i}\cdot \sh{y}{i-1}$ for $i \in [3]$. 
    $(\sh{t}{0}, \sh{t}{1}, \sh{t}{2})$ forms a $(^3_3)$-sharing $\Twoboolshare{t}$. 
    The parties additionally generate a $(^3_3)$-sharing of zero, \ie, $r = \sh{r}{0}+\sh{r}{1}+\sh{r}{2} = 0$. $\party_i$ computes and sends $\sh{z}{i} \leftarrow \sh{t}{i}+\sh{r}{i}$ to $\party_{i+1}$, meanwhile receives $\sh{z}{i-1}$ from $\party_{i-1}$. 
    $\party_i$ sets $\HMboolshare{z}_i \leftarrow (\sh{z}{i}, \sh{z}{i-1})$. 
    A $(^3_3)$-sharing $\Twoboolshare{r}$ of zero can be generated non-interactively using a \textit{PRF-based trick}~\cite{furukawa2017high}: each pair of parties $(\party_{i}, \party_{i-1})$ share a common key $\prfkey{i}$ for a PRF $F:\mathcal{K}\times\mathcal{D} \rightarrow \FF$. Given a session identifier $id \in\mathcal{D}$, $\party_i$ computes $\sh{r}{i} \leftarrow F(\prfkey{i}, id) - F(\prfkey{i+1}, id)$. Clearly, $\sh{r}{0}+\sh{r}{1}+\sh{r}{2} = 0$.
\end{itemize} 

\myindent\textbf{Malicious Security Mechanisms for RSS}. 
\mostree relies on some malicious security mechanisms and functionalities for RSS. 
To start with, we show a \textit{consistent} property for RSS defined as below:  
\begin{definition}[Consistent RSS Sharing \cite{lindell2017framework}]\label{def:correct-RSS}
    Let $(a_1, b_1)$, $(a_2, b_2)$, $(a_3, b_3)$ be the RSS shares held by $\party_0$, $\party_1$, and $\party_2$, and $\party_i$ be corrupted. Then the shares are consistent if and only if $a_{i+1} = b_{i+2}$.     
\end{definition}
One can check that consistency preserves for addition and scalar multiplication.
Consistency also holds for secret-shared multiplication (\ie, $\HMboolshare{x} \cdot \HMboolshare{y}$), despite the fact that a malicious party can add an error (independent of the shared secrets) to resulting RSS sharing;\footnote{
    Suppose $\party_i$ is the corrupted party.  
    $\party_i$ can instead send $z_i \leftarrow t_i + r_i + e$ to $\party_{i+1}$.The parties will share $\HMboolshare{x\cdot y + e}$ instead of $\HMboolshare{x\cdot y}$.   
}
such an attack is known as an \textit{additive attack} in secure computation.  
This is captured by Lemma~\ref{lemma:correctness} (from \cite{lindell2017framework}).  
\begin{lemma}\label{lemma:correctness} 
    If $\HMboolshare{x}$ and $\HMboolshare{y}$ are two consistent RSS sharings and $\HMboolshare{z}$ is generated by executing the multiplication protocol on $\HMboolshare{x}$ and $\HMboolshare{y}$ in the presence of one malicious party, then $\HMboolshare{z}$ is a consistent sharing of either $x \cdot y$ or of some element $z^* \in \FF$. 
\end{lemma}
Existing RSS-based 3PC relies on a \textit{triple verification} \cite{furukawa2017high,mohassel2018aby3} to check the correctness of multiplication.
Given a triple of RSS sharing $(\HMboolshare{a}, \HMboolshare{b}, \HMboolshare{c})$ with $a\cdot b = c$, the parties first open $\HMboolshare{e} = \HMboolshare{x} - \HMboolshare{a}$ and $\HMboolshare{f} = \HMboolshare{y} - \HMboolshare{b}$. 
Then the parties compute $\HMboolshare{w} \leftarrow e\cdot f + f\cdot \HMboolshare{a} + e\cdot \HMboolshare{b} + \HMboolshare{c} - \HMboolshare{z}$ and securely open $\HMboolshare{w}$ to check if $w = 0$.

\mostree uses some assumed ideal functionalities to achieve malicious security, including $\f{rand}$, $\f{open}$, $\f{coin}$, $\f{recon}$, $\f{share}$, $\f{mul}^{\FF}$, and $\f{CheckZero}$ from \cite{chida2018fast,furukawa2017high,lindell2017framework}. 
All these functionalities can be securely computed with malicious security using well-established protocols~\cite{chida2018fast,furukawa2017high,lindell2017framework}.
We also provide the corresponding protocols in Appendix~\ref{appendix:ideal-fun} for completeness. 
\begin{itemize}[leftmargin=10pt]
    \item[$\bullet$] \textbf{$\f{rand}(\FF)$}: sample a random $r \rnd \FF$ and share $\HMboolshare{r}$ between three parties. 

    \item[$\bullet$] $\f{open}(\HMboolshare{x})$: on inputting a \textit{consistent} RSS-sharing $\HMboolshare{x}$, reveal $x$ to all the parties.  

    \item[$\bullet$] \textbf{$\f{coin}(\FF)$}: sample a random $r \rnd \FF$ and output $r$ to three parties.

    \item[$\bullet$] $\f{recon}(\HMboolshare{x}, i)$: on inputting a \textit{consistent} RSS-sharing $\HMboolshare{x}$ and a party index $i$, send $x$ to $\party_i$. 

     \item[$\bullet$] $\f{share}(x, i)$: on inputting a secret $x$ held by $\party_i$, share $\HMboolshare{x}$ between the parites. 
    
    \item[$\bullet$] \textbf{$\f{mul}^{\FF}(\HMboolshare{x}, \HMboolshare{y}, e)$}: take two RSS-sharing $\HMboolshare{x}$ and $\HMboolshare{y}$ for $x,y \in \FF$ and an additive error $e \in 
    \FF$ specificed by the adversary $\ADV$, share $\HMboolshare{x\cdot y + e}$ between three parties. 
    
    \item[$\bullet$] \textbf{$\f{CheckZero}(\HMboolshare{x})$}: take $\HMboolshare{x}$ as input and output $\mathsf{True}$ if $x = 0$ and $\mathsf{False}$ otherwise.
\end{itemize}

In addition, whenever three-party RSS-based secure computation (3PC for short) is used in a black-box manner (\eg, secure comparison and secure MUX in \mostree), we will use a 3PC ideal functionality $\f{3pc}^{\FF}$ directly for simplicity, which ensures privacy and correctness for secure computation over $\FF$ against a malicious adversary.

\myindent\textbf{Distributed Point Function}.
A point function $f_{\alpha, \beta}: \mathcal{D}\rightarrow \mathcal{R}$ outputs $\beta$ only if $x = \alpha$ and outputs $0$ for all $x \in \mathcal{R} \setminus \{\alpha\}$. 
A two-party distributed point function~(DPF) scheme~\cite{gilboa2014distributed,boyle2015function,boyle2016function} can share a point function using two succinct correlated keys~(with size sublinear in $|\mathcal{D}|$). 
Def.~\ref{Def:DPF} shows the formal definition. 
\begin{definition}[Distributed Point Function]  
\label{Def:DPF}
	A two-party DPF scheme ${\rm \Pi_{\rm dpf}} = (\textsf{\rm Gen}, \textsf{\rm Eval}, \textsf{\rm BatchEval})$ consists of three algorithms: 
	\begin{itemize}[leftmargin=10pt]
		\item[$\bullet$] $(\fsskey{0}, \fsskey{1}) \leftarrow \textsf{\rm Gen} (1^{\symsec}, f_{\alpha,\beta})$.  
		Given a security parameter $1^{\symsec}$ and a point function $f_{\alpha, \beta}: \mathcal{D}\rightarrow \mathcal{R}$, outputs a two keys $(\fsskey{0}, \fsskey{1})$, each for one party.
		
		\item[$\bullet$] $\sh{y}{i} \leftarrow \textsf{\rm Eval} (\fsskey{i}, x)$. Given a key $\fsskey{i}$ for party $P_i$~$(i \in \{0, 1\})$, and an evaluation point $x \in \mathcal{D}$, outputs a group element $\sh{y}{i} \in \mathcal{R}$ as the share of $f(x)$ for $P_i$.

        \item[$\bullet$] $\{\sh{y_j}{i}\}_{j \in [L]} \leftarrow \mathsf{BatchEval} (\fsskey{i}, \{x_j\}_{j \in [L]})$. This algorithm performs evaluation over a batch of $L$ inputs $\{x_j\}_{j \in [L]}$, outputs a set of shares $\{\sh{y_j}{i}\}_{j \in [L]}$, where $\sh{y_j}{i} \leftarrow \textsf{\rm Eval} (\fsskey{i}, x_j)$. 
        
	\end{itemize}
\end{definition}
A DPF scheme should ensure secrecy and correctness properties. 
Roughly, secrecy requires that one party cannot learn any more information from its DPF key. 
Correctness requires $\textsf{\rm Eval} (\fsskey{0}, x) + \textsf{\rm Eval} (\fsskey{1}, x) = f_{\alpha,\beta}(x)$ always holds. 
We refer to Appendix~\ref{appendix:security-def} for the formal definition.

\myindent\textbf{Verifiable Distributed Point Function}.
When the party responsible for generating and distributing DPF keys is malicious, it may generate incorrect keys.
To prevent this, verifiable DPFs~(VDPFs)~\cite{de2022lightweight} additionally provide a verifiable property, defined as follows: 
\begin{definition}[Verifiable DPF]  
\label{Def:VDPF}
	A verifiable distributed point function scheme $\mathsf{VDPF} = (\mathsf{Gen}, \mathsf{Eval}, \mathsf{BatchEval}, \mathsf{Verify})$ contains four algorithms. :
	\begin{itemize}[leftmargin=10pt]

        \item[$\bullet$] $\textsf{\rm Gen}$ and $\textsf{\rm Eval}$: Same as the definition in DPF.
        	
        \item[$\bullet$] $(\sh{y_j}{i}, {\pi}_{i}) \leftarrow \mathsf{BatchEval} (\fsskey{i}, \{x_j\}_{j \in [L]})$. This algorithm performs batch evaluation with an additional output $\pi_i$ which is used to verify the correctness of the output. 
        		
        \item[$\bullet$] $\textsf{\rm Accept/Reject} \leftarrow \textsf{\rm Verify}({\pi}_{0}, {\pi}_{1})$. This is a protocol run between the DPF evaluators, which takes the proofs ${\pi}_{0}$ and ${\pi}_{1}$ as the inputs and outputs either $\textsf{\rm Accept}$ or $\textsf{\rm Reject}$. 

	\end{itemize}
\end{definition}

\myindent\textbf{Security Definition}. 
We follow the simulation-based security model in the three-party honest-majority setting~\cite{araki2016high,furukawa2017high}. 
We refer to Appendix~\ref{appendix:security-def} for a formal definition. 


\section{Overview of \mostree} \label{sec:overview}
This section shows the threat model and overview of \mostree.

\subsection{Threat Model}
\myindent\textbf{System Model}. 
\mostree contains three parties: a model owner~(\MO), a feature owner (\FO), and one assistant computing party~(\CP). 
\mostree consists of a one-time setup protocol and an evaluation protocol.  
In the setup protocol, \MO uses RSS sharing to share a tree model among the three parties. 
Whenever \FO wants to perform a PDTE query, it shares its feature vector among the three parties. 
The parties jointly run the evaluation and reconstruct the classification result to \FO, completing the PDTE task. 

\mostree works in the three-party honest-majority setting \cite{furukawa2017high,lindell2017framework,aby3,araki2016high} where at most one party is malicious and other two parties are honest. 
The same assumption is also accepted by many recent privacy-preserving works \cite{le2019two,dauterman2022waldo,mouris2023plasma,aby3} in order to trade a better efficiency that cannot be obtained by two-party protocols.  
\mostree ensures privacy and correctness with abort in the presence of a malicious adversary under this model. 




\myindent\textit{Remark}. 
A secure computation protocol ensures private and correct computation once the inputs are fed into the protocol. 
We do not consider attacks from manipulated inputs or leakage from PDTE protocol output~(\eg, inference attacks and model stealing attacks). 

\subsection{Approach Overview}\label{section:overview:approach}
\mostree first encodes a DT as an array, then transforms DTE to a traversal algorithm over arrays.
\mostree focuses on designing protocols to securely evaluate the DT algorithm over the encoded tree and feature arrays.

\myindent\textbf{Evaluation over Encoded Tree Array}. 
\mostree encodes a DT as a multi-dimensional array $\tree$. 
Fig.~\ref{fig:tree-array} shows the tree encoding method. 
Each array element corresponding to a tree node contains five values: left child index $l$, right child index $r$, threshold value $t$, feature ID $v$, and classification label $c$~(only valid for leaf nodes). 
One can perform DTE over the encoded tree array and a feature vector. 
The evaluation runs at most $d$ iterations, where $d$ is the maximal DT depth.
The evaluation starts from $\tree[0]$. In each iteration, the algorithm fetches $\feature[v]$ and compares it with the current threshold value $t$, and decides to go left or right child according to the comparison result. 
The algorithm terminates and outputs its label $c$ as the classification result when reaching a leaf. 

\begin{figure}
	\centering 
	\includegraphics[height=1.46in]{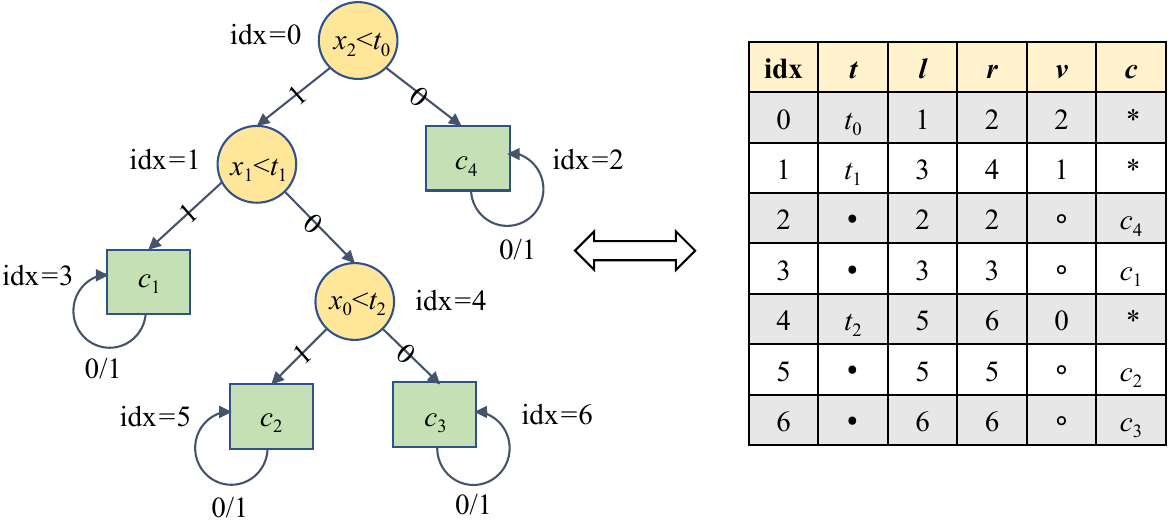} 
    \vspace{-2mm}
	\caption{Encoding Tree as an Tree Array $\tree$ by Breadth-First Search (BFS): $\bullet \in [2^k], \circ \in [n]$ and $* \in [2^k]$ where $k$ is the bit length of single value and $n$ represents feature dimension.
    }
    \label{fig:tree-array}
    \myfigspace
\end{figure}

\myindent\textbf{Private and Correct Tree Evaluation}.  
\mostree aims to evaluate the tree \textit{privately} and \textit{correctly} in the presence of a malicious adversary. 

We address the following security issues with efficient solutions. 
    \ding{182} \textit{Hiding secret values}: a PDTE protocol should hide each element of $\tree$ and $\feature$ as well as all intermediate states/values. 
    \ding{183} \textit{Hiding running time}: a PDTE protocol should hide the decision path length.  
    If \MO learns the number of evaluation rounds, it can infer the accessed path. 
    \ding{184} \textit{Hiding access pattern}: a PDTE protocol should hide access pattern over $\tree$ and $\feature$. Specifically, no party should learn which child and feature are taken during each evaluation round. 
    \ding{185} \textit{Ensuring correct classification}: the \FO must receive a correct classification result if the protocol completes.
To ensure \ding{182}, we choose RSS-based boolean sharing to share all data among three parties, since it matches well with bit-wise computation. 
To ensure \ding{183}, we use the encoding trick from Bai~\etal~\cite{bai2022scalable}. 
As shown in Fig~\ref{fig:tree-array}, the idea is to encode two circles for each leaf node by setting children indexes as the leaf itself, thus the evaluation will be redirected back to itself once reached. The parties always run $d_{\mathsf{pad}} \geq d$ iterations for any query. 
This hides running time meanwhile ensures the correctness of classification.


Achieving \ding{184} under the constraint of sublinear communication and malicious security is challenging. 
What we want is a sublinear-communication oblivious selection~(OS) functionality that allows the parties to obliviously select and share a desired tree node in a secret-shared fashion. 
We propose two efficient OS protocols for RSS sharing. 
However, both protocols are not totally maliciously secure.
They are all vulnerable to additive attacks, which compromise the correctness property. To address this issue, we design a set of lightweight {consistency checks}, exploiting some nice properties of the proposed primitives and reusing existing RSS-based malicious secure mechanisms and functionalities.
Combining them together, \mostree ensures \ding{185}.

\section{The Mostree Protocol} \label{sec:protocol}
\myspace

This section details techniques in \mostree. 
We first propose two oblivious selection~(OS) protocols both with constant online communication. 
Then we combine OS protocols with our tree encoding method and existing 3PC ideal functionalities to design \mostree.

\subsection{Oblivious Selection from Pure RSS}
\mostree uses Oblivious Selection (OS) protocols to perform oblivious node selection. 
Our OS protocols aim to compute functionality $\f{os}$ securely: On inputting an RSS-shared vector $\HMboolshare{\tree}$ and an RSS-shared index $\HMboolshare{\idx}$, receive an error $e \in \FF$ from the adversary, share $\HMboolshare{\tree[\idx] + e}$ (with rerandomization) between the parties. 
Here $\f{os}$ is up to additive attacks. 
Looking ahead, this imperfect $\f{os}$ definition suffices for our purpose and achieves our efficiency goals. 


\myindent\textbf{Oblivious Selection Using Inner-product}. 
We use inner product computation for oblivious selection. 
Specifically, given an RSS-sharing $\HMboolshare{\idx}$ and an RSS-sharing vector $\HMboolshare{\tree}$ and assume the parties can somehow share a unit vector $\HMboolshare{\B{u}}$ such that $\B{u}$ comprises all 0s except for a single 1 at $\idx$, selection can be easily made by computing $\HMboolshare{\tree[\idx]} \leftarrow \HMboolshare{\B{u}\odot \tree}$.

\textit{\underline{Achieve constant resharing communication}}. 
Inner-product computation requires $m$ multiplications. 
Multiplication between two RSS sharings involves a resharing phase that reshares a $(_3^3)$-sharing back to a $(^3_2)$-sharing. Resharing requires communication; thus, trivially invoking $m$ multiplications would incur linear communication. 
We use an optimization trick to reduce the overhead: the parties first sum all intermediate $(^3_3)$-sharings of $m$ multiplication and then perform resharing only once. Now we can achieve constant communication for inner-product computation. 
However, generating $\HMboolshare{\B{u}}$ from $\HMboolshare{\idx}$ requires linear communication: the parties compute $\HMboolshare{\B{u}}$ such that $\B{u}[j] \leftarrow (j==\idx)$ for $j \in [m]$, which requires $m$ invocations of secure equality comparison with linear communication. 

\textit{\underline{Achieve constant online communication}}. 
We use a derandomization technique~\cite{boyle2019secure,bai2022scalable,ji2021uc,attrapadung2021oblivious} to further reduce online communication.  
In particular, suppose the parties have already shared a unit vector $\B{v}$ whose non-zero element appears at a random position $\rdx$. 
When obtained $\idx$, the parties compute $\HMboolshare{\Delta} \leftarrow \HMboolshare{\rdx \oplus \idx}$ and open $\Delta$ by revealing shares to other parties. {Note that the vector length must be power-of-2 to enable this derandomization.}  
Now the parties can define $\HMboolshare{\B{u}}$ such that $\HMboolshare{\B{u}[j]} \leftarrow \HMboolshare{\B{v}[j\oplus\Delta]}$ for $j \in [m]$.
Clearly, $\B{u}[j] = 1$ only for $j = \idx$.  
Then they can use $\HMboolshare{\B{u}}$ and $\HMboolshare{\tree}$ to perform selection, as mentioned above.
Since generating $\HMboolshare{\B{v}}$ can be moved to the offline phase, the online communication is reduced to constant.

\begin{figure}[htbp]
\mysize
	\framebox{\begin{minipage}{0.99\linewidth}
	\small
	{\bf Parameters}:
	An RSS-sharing array $\boolshare{\tree}$ where each element $\tree[j] \in \FF_{{2}^{\ell}}$ for $j \in [m]$; $m = 2^{\ell_m}$ where $\ell_m$ denotes the number of bits of $m$; an RSS-sharing index $\HMboolshare{\idx}$ for $\idx \in [m]$; a PRF $F: \mathcal{K}\times \mathcal{D} \rightarrow \mathcal{R}$ where $\mathcal{K} = \mathcal{D} = \{0,1\}^{\symsec}$ and $\mathcal{R} = \FF_{{2}^{\ell}}$. A common session identifier $\mathsf{sid}$, and each pair of two parties $(\party_{i-1}, \party_{i})$ hold a common PRF key $\prfkey{i} \in \mathcal{K}$ for $i \in [3]$. 

		{\bf [Preprocess]} The parties generate a random unit vector sharing $\HMboolshare{\B{v}}$ with a non-zero index sharing $\HMboolshare{\rdx}$. 
		
        \begin{enumerate}[leftmargin=10pt]
		    \item The parties call $ \HMboolshare{\rdx} \leftarrow \f{rand}({\FF_{2^{\ell_m}}})$.  
            \item The parties call $\f{3pc}^{\FF_2}$ to compute an RSS-shared unit vector $\HMboolshare{\B{v}}$, where $\B{v}[\rdx]=1$ and $\B{v}[j] = 0$ for all $j \ne \rdx$. 
            \item The parties store $(\HMboolshare{\rdx}, \HMboolshare{\B{v}})$ for online computation.  
		\end{enumerate}
		{\bf [Selection]} Upon input $(\mathsf{sid}, \HMboolshare{\tree}, \HMboolshare{\idx})$, do the following:
		\begin{enumerate}[leftmargin=10pt]
		    \item \label{online:step1} Fetch a preprocessed random unit vector RSS-sharing $(\HMboolshare{\rdx}, \HMboolshare{\B{v}})$. 
      
            \item Compute $\HMboolshare{\Delta} \leftarrow \HMboolshare{\rdx} \oplus \HMboolshare{\idx}$ and open $\Delta \leftarrow \f{open}(\HMboolshare{\Delta})$. If the open fails, abort. 

            \item The parties define $\HMboolshare{\B{u}}$ where $\HMboolshare{\B{u}[j]} \leftarrow \HMboolshare{\B{v}[j\oplus \Delta]}$. 

            \item Compute oblivious selection using inner-product as follows: 
            \begin{enumerate}
                \item The parties compute a $(^3_3)$-sharing $\Twoboolshare{t} \leftarrow \sum_{j \in [m]} \Twoboolshare{\tree[j] \cdot \B{u}[j]}$ using $\HMboolshare{\tree}$ and $\HMboolshare{\B{u}}$. 

                \item\label{rss:os:step3.2} For $i \in [3]$: $\party_i$ computes $\sh{r}{i} \leftarrow F(\prfkey{i}, \mathsf{sid}) - F(\prfkey{i-1}, \mathsf{sid})$. $\party_i$ defines $\sh{s}{i} = \sh{t}{i} + \sh{r}{i}$ and sends $\sh{s}{i}$ to $\party_{i+1}$. 

                \item For $i \in [3]$, party $\party_i$ defines $\HMboolshare{s}$ such that $\HMboolshare{s}_i \leftarrow (\sh{s}{i}, \sh{s}{i-1})$. 
            
            \end{enumerate} 
            \item Output $\HMboolshare{s}$.
        \end{enumerate} 
	\end{minipage}}
    \myfigspace
	\caption{Protocol $\protocol{rss\text{-}os}$ with Additive Attacks}
	\label{protocol:RSS-OS}	
\end{figure}

\myindent\textbf{Malicious Security with up to Additive Attacks}. 
To achieve malicious security for the offline phase, we rely on existing malicious secure equality comparison protocol to generate the unit vector $\HMboolshare{\B{v}}$. 
This part is not our focus and we use the existing ideal functionality $\f{3pc}^{\FF_2}$ directly.
We also provide the concrete secure equality comparison protocol $\protocol{eq}$ in Appendix \ref{appendix:ideal-fun} for completeness.


The online phase requires opening $\HMboolshare{\Delta} = \HMboolshare{\rdx} \oplus \HMboolshare{\idx}$, adjusting $\HMboolshare{v}$, and performing an inner product computation. 
However, a malicious party can add errors in the online phase to break correctness. 
First, we assume both $\HMboolshare{\idx}$ and $\HMboolshare{\rdx}$ are \textit{consistent} RSS sharing; we will show this assumption holds when using OS in \mostree. 
Thus opening $\HMboolshare{\Delta}$ can be done \textit{correctly} using $\f{open}$ with malicious security. 
The challenge is to check the correctness of $m$ multiplications.
Note that resharing for multiplication is subject to additive attacks, allowing the adversary to inject an error into the multiplication sharing. 
Existing malicious secure RSS-based 3PC protocols~\cite{furukawa2017high,mohassel2018aby3} perform a triple-based correctness check for \textit{each} multiplication (refer to section~\ref{Pre:3pc-rss}), which incurs linear communication for inner-product. 

Our observation is that we can omit the correctness check for multiplication at this stage, which explains why we formally capture this attack in the definition of $\f{os}$; we rely on a communication-efficient mechanism to detect additive errors at a later stage. 
As a benefit, the resulting OS protocol keeps constant online communication.

\myindent{\textbf{The RSS-based OS Protocol}}. 
The RSS-based OS protocol $\protocol{rss\text{-}os}$ is formally descripted in Fig.~\ref{protocol:RSS-OS}. 
Security of $\protocol{rss\text{-}os}$ is from Theorem~\ref{thm:rss-os-security}, with a proof from Appendix~\ref{proof:rss-sos}. 

\begin{theorem}\label{thm:rss-os-security}
\myspace
$\protocol{rss\text{-}os}$ securely computes $\f{os}$ in the $(\f{rand},$ $ \f{3pc}^{\FF_2},$ $\f{open})$-hybrid model in the presence of a malicious adversary in the three-party honest-majority setting, assuming $F$ is a secure PRF. 
\end{theorem}

\subsection{Oblivious Selection from DPF and RSS}
Protocol $\protocol{rss\text{-}os}$ enjoys constant online communication but linear offline communication. 
We propose another OS protocol with constant online communication and sublinear offline communication. 

\myindent\textbf{A Semi-honest DPF-based OS Protocol}.
The idea is similar to $\protocol{rss\text{-}os}$. 
The difference is we apply DPFs over RSS sharings to perform inner-product computation. 
We first review how to construct a semi-honest OS protocol, then securely enhance it to realize $\f{os}$.

\begin{figure}[t!]
	\centering 
	\includegraphics[height=2in]{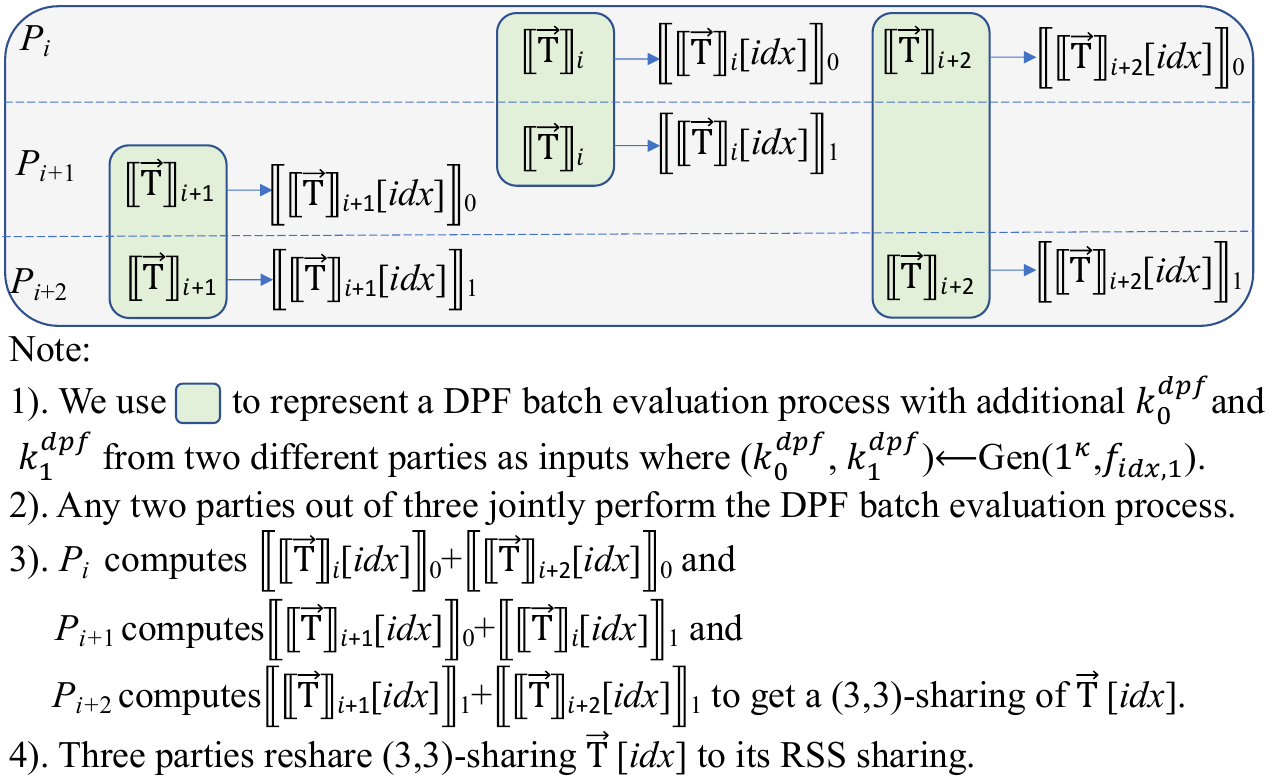} 
       
        \myfigspace
	\caption{OS from DPF and RSS
 }
    \label{fig:osdpf}

\end{figure}

\textit{\underline{Apply DPF over RSS sharings}}. 
Recall that for an RSS-shared tree array $\HMboolshare{\tree}$, $\party_i$ and $\party_{i+1}$ hold a common share vector $\sh{\tree}{i}$. 
$\party_i$ and $\party_{i+1}$ can apply a DPF over the common $\sh{\tree}{i}$ to perform OS operations. 
Specifically, a trusted dealer generates a pair of DPF keys $(\fsskey{0},\fsskey{1}) \leftarrow \textsf{Gen}(1^{\symsec}, f_{\idx, 1})$ and sends $\fsskey{b}$ to $\party_{i+b}$ for $b \in \{0,1\}$, respectively. 
Each party locally expands its DPF key over the domain $[m]$. 
By doing this, $\party_i$ and $\party_{i+1}$ share a vector $\Twoboolshare{\B{u}}$ in $(^2_2)$-sharing, where $\B{u}$ comprises all 0s except for a single $1$ appearing at coordinate $\idx$. 
Each party locally computes inner-product 
$\Twoboolshare{\sh{\tree}{i}[\idx]} \leftarrow \sum_{j \in [m]}\Twoboolshare{\B{u}[j]} \cdot \sh{\tree}{i}[j],$
a $(^2_2)$-sharing of $\sh{\tree}{i}[\idx]$ between $\party_i$ and $\party_{i+1}$. 
The above process is repeatedly run for any two parties out of three.
At the end of the three-round execution, each party will hold two shares of $(^2_2)$-sharing~(since each party will run oblivious selection twice with the other two parties). 
After that, the parties locally sum up two local shares of $(^2_2)$-sharing. 
In this manner, three parties jointly produce a $(^3_3)$-sharing of ${\tree[\idx]}$, which can be reshared back to an RSS-sharing $\HMboolshare{\tree}$ using the PRF-based re-sharing trick~\cite{furukawa2017high} (also refer to Section~\ref{Pre:3pc-rss} and Fig.~\ref{protocol:RSS-OS}). Fig.~\ref{fig:osdpf} sketches the DPF-based OS protocol.

To remove the trusted party, we let the non-involved party $\party_{i+2}$ generate the DPF keys. 
However, we cannot let $\party_{i+2}$ know $\idx$ as this will violate privacy. 
We apply the same derandomization technique used in $\protocol{rss\text{-}os}$ to resolve the issue.  
Specifically, in the offline phase, $\party_{i+2}$ generates a pair of DPF keys for a point function  $f_{\rdx,1}$ with a randomly chosen index $\rdx$, and $\party_{i+2}$ shares the DPF keys as well as $\rdx$ (using $(^2_2)$-sharing) between $\party_i$ and $\party_{i+1}$. 
In the online phase, $\party_i$ and $\party_{i+1}$ each expands its DPF key locally to share a $(^2_2)$-sharing of a unit vector $\Twoboolshare{\B{v}}$. 
$\party_i$ and $\party_{i+1}$ then open $\Delta = \rdx \oplus \idx$ and compute $\HMboolshare{\B{u}}$ such that $\B{u}[j] \leftarrow \B{v}[j \oplus \Delta]$ for $j \in [m]$. 
With $\Twoboolshare{\B{u}}$ and $\sh{\tree}{i}$, $\party_i$ and $\party_{i+1}$ can locally compute and share $\Twoboolshare{\tree_i[\idx]}$ in $(^2_2)$-sharing. 
Note that the PDF keys are of size $O(\symsec\log m)$ where $\symsec$ is the security parameter, and the key generation can be moved to the offline phase.\footnote{The concrete offline efficiency is poor for small trees. In this case, $\party_{i+2}$ directly shares the unit vector $\B{v}$ between $\party_i$ and $\party_{i+1}$ instead of distributing a pair of DPF keys.}
Efficiency-wise, this requires constant online communication and sublinear offline communication.

\myindent\textbf{Attacks}.
The DPF-based OS protocol involves three interactive parts: 
1) DPF key generation and distribution; 2) opening $\Delta$; and 3) DPF-based local evaluation and resharing. 
Now we discuss malicious attacks for each part.

Firstly, we discuss attacks from a corrupted key generator $\party_{i+2}$. 
There are two possible attacks: 
1) {\textit{Incorrect DPF keys}}. Instead of generating a pair of well-formed DPF keys, the corrupted party may generate the keys for a point function $f_{\rdx, s}$ with $s \ne 1$; Indeed, the corrupted party may take arbitrary key generation strategies.  
2) {\textit{Incorrect index sharing}}. Even the DPF keys for $f_{\rdx, 1}$ are correctly generated and shared, the malicious $\party_{i+2}$ may share an inconsistent index $\rdx^* \ne \rdx$ between $\party_i$ and $\party_{i+1}$. 
As a consequence, this attack results in an incorrect opening of $\Delta \leftarrow \rdx \oplus \idx \oplus e$ where the error $e = \rdx \oplus \rdx^*$; looking ahead, this corresponds to an incorrect node being selected in \mostree. 
Secondly, we notice that even if the DPF keys are correctly generated and the index $\rdx$ is correctly shared, how to correctly reconstruct $\Delta$ to $\party_i$ and $\party_{i+1}$ is still a question. 
If only $\party_i$ and $\party_{i+1}$ are involved in the reconstruction of $\Delta$, it is impossible to ensure correctness because an adversary who corrupts either $\party_i$ or $\party_{i+1}$ can always add errors during the reconstruction process.  
Lastly, a corrupted party may maliciously add errors before resharing the $(^3_3)$-shared selected secret, resulting in an additive attack. We note that the last attack is allowed by $\f{os}$.

\myindent\textbf{Defences}. 
The above attacks only compromise correctness, not privacy. 
We propose a set of consistency checks to ensure correctness in the above protocol.  
Now we present intuitions, and the formal description is shown by protocol $\protocol{dpf\text{-}os}$ in Fig.~\ref{proto:dpf-os}. 

\begin{figure}[htbp]
\myspace
\mysize
	\framebox{\begin{minipage}{0.99\linewidth}
	\small
	{\bf Parameters}: 
	    An RSS-sharing for array $\boolshare{\tree}$ where each element $\tree[j] \in \FF_{{2}^{\ell}}$ for $j \in [m]$;
	    $m = 2^{\ell_m}$ where $\ell_m$ denotes the number of bits of $m$; 
		an RSS-sharing index $\HMboolshare{\idx}$ for $\idx \in [m]$;
		point function $f_{\alpha, \beta} : [m] \rightarrow \FF_{2^{\ell}}$;
		each pair of two parties $(\party_i, \party_{i+1})$ hold a common key ${\prfkey{i}} \in \{0,1\}^{\symsec}$ for $i \in [3]$ for a PRF $F: \mathcal{K}\times \mathcal{D} \rightarrow \mathcal{R}$ where $\mathcal{K} = \mathcal{D} = \{0,1\}^{\symsec}$ and $\mathcal{R} = \FF_{{2}^{\ell}}$. A common session identifier $\mathsf{sid}$.  
	

		{\bf [Preprocess]} The parties run the following protocol to generate sufficiently many DPF keys. 
		\begin{enumerate}[leftmargin=10pt]
		    \item $P_{i+2}$ samples a random value $\rdx \rnd \ZZ_{m}$ and locally computes a pair of DPF keys $(\fsskey{0},\fsskey{1}) \leftarrow \mathsf{VDPF}.\mathsf{Gen}(1^{\kappa}, f_{\rdx,1})$.  
		    $\party_{i+2}$ samples $\sh{\rdx}{0}$ and $\sh{\rdx}{1}$ such that $\sh{\rdx}{0}+\sh{\rdx}{1} = \rdx$ over $\ZZ_m$. 
		    $\party_{i+2}$ sends $(\fsskey{0}, \sh{\rdx}{0})$ and $(\fsskey{1}, \sh{\rdx}{1})$ to $\party_{i}$ and $\party_{i+1}$, respectively. 

            \item\label{dpf-os:pre:step2} $\party_{i}$ and $\party_{i+1}$ run the DPF key verification protocol from \cite{de2022lightweight} to check the well-formness of DPF keys. If the check fails, abort. 

            \item\label{dpf-os:pre:step3} $\party_{i}$ and $\party_{i+1}$ each expand its DPF key over domain $[m]$ to jointly produce a shared vector $\Twoboolshare{\B{v}}$ in the $(^2_2)$-sharing. 
            For $b \in \{0,1\}$, $\party_{i+b}$ locally computes: 
            $$
            {\centering 
            \Twoboolshare{t}_{b} \leftarrow \sum_{j \in [m]}\Twoboolshare{\B{v}[j]}_b, \Twoboolshare{s}_{b} \leftarrow \Twoboolshare{\rdx}_b - \sum_{j \in [m]}j \cdot \Twoboolshare{\B{v}[j]}_b.
            }
            $$
            $\party_{i}$ and $\party_{i+1}$ open $\Twoboolshare{t}$ and $\Twoboolshare{s}$ and check if $t = 1$ and $s = 0$. If the check fails, abort.

            \item $\party_{i+2}$ stores $(\rdx, \Twoboolshare{\rdx}_0, \Twoboolshare{\rdx}_1)$, $\party_i$ stores $(\fsskey{0}, \Twoboolshare{\rdx}_0)$, and $\party_{i+1}$ stores $(\fsskey{1}, \Twoboolshare{\rdx}_1)$. 
            
		\end{enumerate}
		{\bf [Selection]} Upon input $(\mathsf{sid}, \HMboolshare{\tree}, \HMboolshare{\idx})$, do:
		\begin{enumerate}[leftmargin=10pt]
		    \item \label{online:step1} For $i \in [3]$: 
		    \begin{enumerate}[leftmargin=10pt]
                \item For $b \in \{0,1\}$, $\party_{i+b}$ fetches a DPF key with the index share $(\fsskey{b}, \Twoboolshare{\rdx}_b)$ distributed by $\party_{i+2}$ in the offline phase. 
                Then $\party_{i}$ sets $\HMboolshare{\rdx}_i = (0, \sh{\rdx}{0})$, $\party_{i+1}$ sets $\HMboolshare{\rdx}_{i+1} = (\sh{\rdx}{1}, 0)$, and $\party_{i+2}$ sets $\HMboolshare{\rdx}_{i+2} = (\sh{\rdx}{0}, \sh{\rdx}{1})$. 
		        
		        \item Three parties compute $\HMboolshare{\Delta} \leftarrow \HMboolshare{\rdx} \oplus \HMboolshare{\idx}$ and reconstruct $\Delta$ \textit{only} to $\party_{i+1}$ and $\party_{i+2}$ using $\f{recon}$.  
                If reconstruction fails, abort.   
		        
		        \item $\party_{i}$ and $\party_{i+1}$ share a $(^2_2)$-sharing $\Twoboolshare{\sh{\tree}{i}[\idx]} \leftarrow \sum_{j \in [m]} \sh{\tree}{i}[j] \cdot \Twoboolshare{\B{v}[j \oplus \Delta]}$; $\party_{i+b}$ holds $\Twoboolshare{\sh{\tree}{i}[\idx]}_b$ for $b \in \{0,1\}$.  
		    \end{enumerate} 
		    
		    \item Now $\tree_{i}[\idx]$ is shared between $(\party_{i}, \party_{i+1})$ in $(^2_2)$-sharing for $i \in [3]$(each party will hold two $(^2_2)$-shares after the DPF-based evaluation).     
            Each party locally sums up its two shares of $(^2_2)$-sharing to generate a $(^3_3)$-sharing $\Twoboolshare{\tree[\idx]}$.  
		    
		    \item For $i \in [3]$, $\party_i$ computes $\sh{z}{i} \leftarrow \Twoboolshare{\tree[\idx]}_i + F(\prfkey{i}, \mathsf{sid}) - F(\prfkey{i-1}, \mathsf{sid})$ and sends $\sh{z}{i}$ to $\party_{i+1}$. $\party_i$ receives $\sh{z}{i-1}$ from $\party_{i-1}$.  
      
            \item For $i \in [3]$, $\party_i$ defines $\HMboolshare{z}_i \leftarrow (\sh{z}{i}, \sh{z}{i-1})$. 
		\end{enumerate}
	\end{minipage}}
    \myfigspace
	\caption{Protocol \protocol{dpf\text{-}os} from DPF and RSS}\label{proto:dpf-os} 
	\label{protocol:OS}	
\end{figure}

\textit{\underline{Detect malicious DPF keys}}. 
We use a VDPF scheme to prevent a corrupted DPF key generation party from distributing incorrect DPF keys.  
A VDPF scheme additionally allows two key DPF receivers to run an efficient check protocol to test whether the DPF keys are correctly correlated without leaking any information other than the validity of the keys. 
We use an existing VDPF construction~\cite{de2022lightweight} to perform the check for DPF keys. 

Unfortunately, the check method from \cite{de2022lightweight} only ensures the DPF keys correspond to a point function $f_{\alpha, \beta}$ for arbitrary $\beta$; it does not check $\beta = 1$. 
Besides, for our purpose, $\party_i$ and $\party_{i+1}$ must check whether the $(^2_2)$-sharing $\Twoboolshare{\rdx}$ is consistent with the point function $f_{\alpha, \beta}$ shared by the DPF keys~(\ie, ensuring $\rdx = \alpha$). 
To this, we additionally propose the following lightweight checks.
First, to check $\beta = 1$, $\party_i$ and $\party_{i+1}$ locally expand their DPF keys to produce a shared a $(^2_2)$-sharing vector $\Twoboolshare{\B{v}}$, and they check the sum of all the entries is equal to 1. 
Second, to check the shared index $\rdx$ is equal to $\alpha$, the checking parties jointly compute $s = \rdx - \sum_{j \in [m]} j\cdot \B{v}[j]$ in a shared fashion, open $s$, and check if $s = 0$.\footnote
{
When either $\party_i$ or $\party_{i+1}$ is corrupted, the party $\party_{i+2}$ is assumed to be honest in honest-majority setting. 
In view of this, the check protocol has no need to ensure correctness as incorrect opening (either from a corrupted $\party_i$ or $\party_{i+1}$) only makes the protocol abort. 
} 
With these additional checks, we can ensure the correctness of DPF keys and the consistency of the shared index.   
The detailed description can be found from $\protocol{dpf\text{-}os}$ (Step~\ref{dpf-os:pre:step2} - \ref{dpf-os:pre:step3}, preprocess protocol). 

{
    

\textit{\underline{Reconstruct $\Delta$ correctly}}. 
To open $\Delta$ correctly, we observe that any $(^2_2)$-sharing $\Twoboolshare{x}$ can be converted to an RSS sharing $\HMboolshare{x}$ \textit{without interaction}, where $\Twoboolshare{x}$ is distributed by the dealer $\party_{i+2}$ and is shared between $\party_{i}$ and $\party_{i+1}$. 
Our insight is that the dealer $\party_{i+2}$ knows $x$ as well as the shares $\Twoboolshare{x}_0$ and $\Twoboolshare{x}_1$. 
Therefore, the parties can define an RSS sharing $\HMboolshare{x}$ from $\Twoboolshare{x} = \{\Twoboolshare{x}_0, \Twoboolshare{x}_1\}$ \textit{non-interactively} as: 
$$ 
\HMboolshare{x}_{i} = (0, \Twoboolshare{x}_0), \; \HMboolshare{x}_{i+1} =  (\Twoboolshare{x}_1, 0), \; \HMboolshare{x}_{i+2} = (\Twoboolshare{x}_0, \Twoboolshare{x}_1). 
$$
Using this trick, the parties can define a \textit{consistent} RSS-sharing $\HMboolshare{\rdx}$ from $\Twoboolshare{\rdx}$. 
Note $\HMboolshare{\idx}$ is a consistent RSS sharing and $\HMboolshare{\Delta} = \HMboolshare{\rdx} \oplus \HMboolshare{\idx}$, then $\HMboolshare{\Delta}$ is also consistent (see Def.~\ref{def:correct-RSS} and its following explaination).  
Since $\Delta$ is a consistent RSS sharing, the parties can correctly reconstruct $\Delta$ to $\party_i$ and $\party_{i+1}$ using $\f{recon}$.
Note that $\Delta$ is only reconstructed to $\party_{i}$ and $\party_{i+1}$, $\party_{i+2}$ cannot learn it; otherwise $\party_{i+2}$ can learn $\idx = \Delta \oplus \rdx$.  

After applying the above consistency checks, a corrupted party is only limited to adding errors in the resharing phase.


\myindent\textbf{DPF-based OS Protocol}.
The protocol is formally described in Fig.~\ref{proto:dpf-os}. 
We note that $\protocol{dpf\text{-}os}$ is subject to an additive attack in the resharing phase; we will show how to handle additive errors at a later stage. 
Theorem~\ref{thm:sos-security} shows the security of $\protocol{dpf\text{-}os}$, with the proof from Appendix~\ref{proof:sos}.
\begin{theorem}\label{thm:sos-security}
\myspace
\protocol{dpf\text{-}os} securely computes $\f{os}$ in the $\f{recon}$-hybrid model in the presence of a malicious adversary in the three-party honest-majority setting, assuming $F$ is a secure PRF.
\end{theorem}

\subsection{The \mostree PDTE Protocol}\label{section:mostree:protocol}

\myindent\textbf{Detecting Errors using MACs}.
\mostree uses $\f{os}$ for oblivious node selection, but $\f{os}$ is subject to additive errors. 
We use SPDZ-like MACs to detect errors.


\textit{\underline{SPDZ-like MACs}}.
An SPDZ-like MAC $\sigma(\alpha, x)$ is usually defined as $\sigma(\alpha, x) = \alpha \cdot x$ over a finite field $\FF$, where $x$ is the value to be authenticated and $\alpha$ is the MAC key. 
These MACs are additively homomorphic: $\alpha \cdot (x + y) = \alpha\cdot x + \alpha \cdot y$. 
In the following, we will drop $\alpha$ and instead use $\sigma(x)$ or $\sigma_x$ to denote the MAC for $x$ when the context is clear. 
A MAC $\sigma(x)$ is also shared along with its authenticated secret $x$. 
Secure computation is performed both for $x$ and $\sigma(x)$, and the protocol aborts if the output $x$ and the MAC tag $\sigma_x$ not satisfying $\alpha \cdot x = \sigma_x$.
In order to achieve overwhelming detection probability, the field $\FF$ must be large enough, which is vital to ensure overwhelming detection probability.

\textit{\underline{MACs over $\FF_{2^{\ell}}$ with low overhead}}. 
We use MACs over $\FF_{2^{\ell}}$ to authenticate $\ell$-bit secrets as a whole.
We exploit the fact that $\ZZ_2^{\ell}$ is compatible with $\FF_{2^{\ell}}$ over addition~(\ie, bit-wise XOR). 
Therefore, a secret shared over $\ZZ_2^{\ell}$ can be converted to a secret over $\FF_{2^{\ell}}$ for free. 
We note that $\FF_{2^{\ell}}$ is incompatible with bit-wise multiplication. 
Nevertheless, these MACs are only used for detecting errors from oblivious selection rather than the whole computation.  
Another benefit is that $\ell > \statsec$ for real-world decision-tree applications where $\statsec$ is a statistical security parameter. Thus $\FF_{2^{\ell}}$ is large enough for error detection. 
This means the MACs are at the same size as the authenticated secrets, incurring only constant overhead. 
Notably, our method does not require any expensive share conversion. 

\begin{figure}[htbp]
\mysize
	\framebox{\begin{minipage}{0.95\linewidth}
		{\bf Parameters}: 
		Three parties denoted as $\party_0$, $\party_1$ and $\party_2$; statistical security parameter $\statsec$; 
		a finite field $\FF_{2^{\ell}}$ where $\ell \gg \statsec$; number $m \in \ZZ$ denotes the number of RSS sharings to be checked. 
		
		\smallskip 
		{\bf [Check]} On inputting $m$ RSS sharings $\{x_j\}_{j \in [m]}$, MAC values $\{\sigma(x_j)\}_{j \in [m]}$ and MAC key sharing $\HMboolshare{\alpha}$, outputs $\mathsf{True}$ if MAC check passed and $\mathsf{False}$ otherwise. 
		
		\begin{enumerate}[leftmargin=10pt]
    		\item The parties call $\HMboolshare{r} \leftarrow \f{rand}({\FF_{2^{\ell}}})$ and $\HMboolshare{\sigma(r)} \leftarrow \f{mul}^{\FF_{2^{\ell}}}(\HMboolshare{r}, \HMboolshare{\alpha})$.
    		
    		\item The parties call $\f{coin}(\FF_{2^{\ell}}^m)$ to receive random elements $\rho_1, \rho_2, \cdots, \rho_m \in \FF_{2^{\ell}} \setminus \{0\}$. 
    		
    		\item The parties locally compute $\HMboolshare{v} \leftarrow \HMboolshare{r} + \sum_{j \in [m]} \rho_j \cdot \HMboolshare{x_j}$ and $\HMboolshare{w} \leftarrow \HMboolshare{\sigma(r)} + \sum_{j \in [m]} \rho_j \cdot \HMboolshare{\sigma(x_j)}$.  
    
    		\item Securely open $\HMboolshare{v}$ via $\f{open}$. Abort if the open fails.
    		
    		\item Call $\f{CheckZero}(\HMboolshare{w} - v\cdot \HMboolshare{\alpha})$ and abort if receives $\mathsf{False}$.
		
		\end{enumerate}
	\end{minipage}}
\myfigspace
	\caption{Protocol for Batch MAC Check~\protocol{MacCheck}}
	\label{protocol:batch-mac-checking}
\end{figure}
\myspace

\mostree authenticates the shared tree array $\HMboolshare{\tree}$ by computing their MACs (also shared) in the setup phase. 
In the online phase, the parties run $\f{os}$ over both the shared tree array and its MAC array.
The intuition is that any introduced errors will break the relationship between the shared data and its MAC, which can be detected by a MAC check protocol $\protocol{MacCheck}$ in Fig.~\ref{protocol:batch-mac-checking}. 
Such a batch MAC check technique is previously used in~\cite{chida2018fast,lindell2017framework, hoang2020macao}, which detects additive errors with probability $1 - O(1/|\FF|)$, which is closed to $1$ except with negligible probability, for large enough field $\FF$.

\myindent\textbf{The \mostree Protocol}. 
Combining tree encoding, $\f{os}$, MAC over $\FF_{2^{\ell}}$ and RSS-based secure computation, we propose \mostree formally described in Fig.~\ref{protocol:pdte}.
\mostree intends to compute a PDTE functionality $\f{pdte}$: On inputting a feature vector $\feature$ from \MO and a decision tree $\tree$ from \MO, outputs a classification result to \FO.  
\mostree contains a one-time offline setup protocol and an online evaluation protocol. 

\myspace
\begin{figure}[htbp]
\mysize
	\framebox{\begin{minipage}{0.95\linewidth}
	{\bf Parameters}: 
	    Three parties denoted as $\party_0$, $\party_1$ and $\party_2$; statistical security parameter $\statsec$; 
	    $\FF_2$ denotes the binary field for boolean sharing; 
	    $m$ denotes the number of tree nodes;
	    $\ell_m$ denotes the minimal value such that $m \le 2^{\ell_m}$;
	    $n$ denotes the dimension of feature vectors;
	    $k$ denotes the bit-length of values~(\eg, left and right children index, threshold value, and classification result);
	    $\ell$ denotes the bit-length of tree nodes, \ie, $\ell = 4k+n$;
		a finite field $\FF_{2^{\ell}}$ where $\ell \gg \statsec$. 
	    

		{\bf [Setup]} Upon receiving a DT $\rawtree$, the setup protocol outputs $(\HMboolshare{\tree}, \HMboolshare{{\treemac}})$ where $\HMboolshare{\tree}$ is the RSS-sharing array for the tree and $\HMboolshare{{\treemac}}$ is the MAC array for $\HMboolshare{\tree}$ such that $\treemac = \sigma(\tree)$.
		
		\begin{enumerate}[leftmargin=10pt]
		    \item \MO encodes $\rawtree$ as an array $\tree$. The parties call $\f{share}$ in which \MO inputs $\tree$. As a result, $\HMboolshare{\tree}$ is shared between the parties. 
		    
		    \item \label{pdte:setup:2} The parties call $\HMboolshare{\mackey} \leftarrow \f{rand}({\FF_{2^{\ell}}})$ to share a MAC key $\mackey \in {\FF_{2^{\ell}}}$.
      
            \item \label{pdte:setup:3} Compute MACs: $\HMboolshare{\treemac[j]} \leftarrow \f{mul}^{\FF_{2^{\ell}}}(\HMboolshare{\mackey}, \HMboolshare{\tree[j]})$ for $j \in [m]$. 
		    
		    \item \label{pdte:setup:4} The parties run $\protocol{MacCheck}(\HMboolshare{\tree},\HMboolshare{\treemac}$. If the check fails, abort.  
		    
		    \item Output $(\HMboolshare{\tree}, \HMboolshare{\treemac})$. 
		\end{enumerate}
	
		{\bf [Evaluation]} Upon receiving $\feature$ from \FO, do:  
		\begin{enumerate}[leftmargin=10pt]
            \item \FO shares $\HMboolshare{\feature}$ via calling $\f{share}$. 
		    
		    \item Initialize $\HMboolshare{result} \leftarrow \perp$.
            \item Parse $\HMboolshare{t}||\HMboolshare{l}||\HMboolshare{r}||\HMboolshare{\B{v}}||\HMboolshare{c} \leftarrow \HMboolshare{\tree[0]}$. 
		    
		    \item \label{step:3} For $j \in [d_{\rm pad}]$: // step 1) to 3) are executed by calling $\f{3pc}^{\FF_2}$. 
		    \begin{enumerate}
		        \item $\HMboolshare{x} \leftarrow \sum_{j \in [n]} \HMboolshare{\feature[j]} \cdot \HMboolshare{\B{v}[j]}$.
		        
		        \item $\HMboolshare{b} \leftarrow \HMboolshare{x} < \HMboolshare{t}$. 
		        
		        \item $\HMboolshare{\idx} \leftarrow \HMboolshare{r} \oplus \HMboolshare{b}\cdot (\HMboolshare{l} \oplus \HMboolshare{r})$.
		        
		        \item $\HMboolshare{\tree[\idx]}||\HMboolshare{\treemac[\idx]}$ $\leftarrow$ $\f{os}(\HMboolshare{\idx}, \HMboolshare{\tree}||\HMboolshare{\treemac})$.  
		        
		        \item Parse $\HMboolshare{t}||\HMboolshare{l}||\HMboolshare{r}||\HMboolshare{\B{v}}||\HMboolshare{c} \leftarrow \HMboolshare{\tree[\idx]}$.
		        
		        \item Update $\HMboolshare{result} \leftarrow \HMboolshare{c}$. 
		    \end{enumerate}
		    
		    \item Use $\protocol{MacCheck}$ to check $d_{pad}$ pairs of RSS sharings from $\f{os}$, abort if the check fails.
      
		    \item Call $\f{recon}(\HMboolshare{result})$ to reconstruct ${result}$ to \FO. 
		\end{enumerate}
	\end{minipage}}
\myfigspace
	\caption{The \mostree Protocol $\protocol{pdte}$ }
	\label{protocol:pdte}	
\end{figure}

\textit{\underline{Setup phase}}. 
The setup protocol requires linear communication in the tree size. Setup is only run once thus the overhead can be amortized across subsequent queries. 

As a requirement of $\f{os}$, we need the dimension of $\tree$ to be a power of two~(\ie, $m = 2^{\ell_m}$ for some $\ell_m$). 
If this is not the case, the tree holder can pad the array before sharing the encoded tree. 
In particular, the holder computes the minimal $\ell_m$ such that $m \le 2^{\ell_m}$ and allocates an array of size $2^{\ell_m}$. 
The holder randomly assigns $m$ nodes within the power-of-two-length array and modifies each node's left and right child correspondingly. 
Overall, this padding, at most, doubles the storage overhead of the non-padded version.

\mostree relies on existing ideal functionality $\f{mul}^{\FF_{2^{\ell}}}$ for MAC generation. 
Concretely, the parties first invoke $\f{rand}$ to obtain an RSS-sharing $\HMboolshare{\alpha}$ for a random MAC key $\alpha \rnd \FF_{2^{\ell}}$ (Step~\ref{pdte:setup:2}, Setup). 
Then they call $\f{mul}^{\FF_{2^{\ell}}}$ $m$ times to compute an array sharings $\HMboolshare{\treemac}$ for MACs of $\HMboolshare{\tree}$, where $\treemac[j] = \alpha \cdot \tree[j]$ is defined over $\FF_{2^{\ell}}$ for $j \in [m]$ (Step~\ref{pdte:setup:3}, Setup). 
Note that $\f{mul}^{\FF_{2^{\ell}}}$ also suffers from additive attacks. 
To check whether these MACs are correctly generated, the parties can run batch MAC check over $(\HMboolshare{\tree}, \HMboolshare{\treemac})$ at the end of the setup protocol (Step~\ref{pdte:setup:4}, Setup); {this means we can use \protocol{MacCheck} to uniformly handle leakage from the setup protocol and $\f{os}$.}

\textit{\underline{Evaluation phase}}. 
The evaluation protocol, on inputting $\HMboolshare{\tree}$, $\HMboolshare{\treemac}$ and an RSS-sharing feature vector $\HMboolshare{\feature}$, outputs an RSS-sharing $\HMboolshare{result}$, where $\mathit{result}$ denotes the classification result and will be reconstructed to \FO. 
We stress that feature selection must also be done obliviously. 
Certainly, we can resort to $\f{os}$ again, but it turns out that this approach is overkill in most cases because feature vectors are usually with low-dimension. 
We instead use a much simpler approach. 
Specifically, we modify the tree encoding method by replacing each feature index $v \in \ZZ_{n}$ to a length-$n$ unit bit vector $\B{v} \in \ZZ_2^{n}$, where $\B{v}$ comprises all 0s except for a single $1$ appearing at coordinate $v$.
In this manner, the parties simply run RSS-based inner-product computation for oblivious feature selection. {Here the triple-verification method from \cite{furukawa2017high} is used to ensure the correctness of each multiplication, incurring $O(n)$ communication.} 

The evaluation protocol contains $d_{pad} > d$ iterations. In each iteration, the parties obliviously select the desired feature value, make a secure comparison, and decide on the next node for evaluation. 
All non-RAM computation (\eg, secure comparison) is performed in a secret-shared fashion using 3PC ideal functionality $\f{3pc}^{\FF_2}$, in which privacy and correctness are guaranteed in the presence of a malicious adversary. 
After that, the parties call $\f{os}(\HMboolshare{\tree}||\HMboolshare{\treemac}, \HMboolshare{\idx})$; here we abuse the notation by calling $\f{os}$ over  $\HMboolshare{\tree}$ and $\HMboolshare{\treemac}$ simoustionusly. 
Note that $\f{os}$ is subject to additive attacks, the parties must run MAC check to detect any error for each selection. 
\mostree delays all these MAC checks in a batch, before reconstructing $\HMboolshare{result}$ to the \FO.  

\myindent\textbf{Complexity}. 
The online communication complexity of $\protocol{pdte}$ is sublinear in $m$ because either $\protocol{rss\text{-}os}$ or $\protocol{dpf\text{-}os}$ only requires constant online communication, while the computation complexity is linear since each party needs to perform a linear scan over its local share. 
The linear scan is extremely cheap (\ie, bit-wise computation). Thus, existing hardware acceleration can be applied to optimize its performance.
As for offline communication, the setup protocol \protocol{setup} requires linear communication but is only {invoked} once in the initialization phase, which means all initialized RSS sharings can be reused to perform subsequent evaluation queries, among which the linear communication can be amortized.
For OS setup protocols, $\protocol{rss\text{-}os}$ still requires linear communication while $\protocol{dpf\text{-}os}$ enjoys sublinear offline communication.

\myindent\textbf{Security}. 
We design \mostree in a modular manner, making it easier to analyze security in the hybrid model. 
Formally, we show the security of $\protocol{pdte}$ in Theorem~\ref{thm:pdte-security} and prove it in Appendix~\ref{proof:pdte}. 

\begin{theorem}\label{thm:pdte-security}
\myspace
Protocol \protocol{pdte} securely computes $\f{pdte}$ in the $(\f{3pc}^{\FF_2}$, $\f{os}$, $\f{coin}$, $\f{rand}$, $\f{open}$  $\f{mul}^{\FF_{2^{\ell}}}$, $\f{CheckZero})$-hybrid model for $\ell > \statsec$ in the present of the malicious adversary in the 3PC honest-majority setting.
\end{theorem}

\myspace
\begin{table}
\footnotesize
\renewcommand{\arraystretch}{1}
\caption{Parameters of Datasets}\label{tab::paramters}
\vspace{-2mm}
\centering
	\begin{tabular}{c|c|c|c|c}
	\hline
	\textbf{Dataset} & \textbf{Depth} $d$ & \textbf{Features} $n$ & $\#$(\textbf{Nodes}) $m$ & $\#$(\textbf{Padded nodes}) $m'$\\
	\hline
    wine & 5 & 7 & 23 & 32\\
	breast & 7 & 12 & 43 & 64\\
    digits & 15 & 47 & 337 & 512\\
	spambase & 17 & 57 & 171 & 256\\
	diabetes & 28 & 10 & 787 & 1024\\
	Boston & 30 & 13 & 851 & 1024\\
	MNIST & 20 & 784 & 4179 & 8192\\
	\hline
    \end{tabular}
\end{table}
\myspace
\section{Experiment}\label{sec:experiment}
\myspace
This section reports the implementation and performance of \mostree.

\subsection{Implementation and Experiment Details}
We evaluate the performance of \mostree using ABY$^3$~\cite{aby3} in C++.
Our implementation is available at \url{https://github.com/Jbai795/Mostree-pub}.
We test on 7 representative datasets from the UCI repository~({\url{https://archive.ics.uci.edu/ml}}) as listed in Table~\ref{tab::paramters}.
The trees we trained vary in depth and size. Among them, \textit{wine, breast} are small example trees while \textit{Boston} is a deep-but-sparse tree and \textit{MNIST} is a density tree with a high-dimensional feature vector.

We run benchmarks on a desktop PC equipped with Intel(R) Core i9-9900 CPU at 3.10~GHz × 16 running Ubuntu 20.04 LTS with 32~GB memory. 
We use Linux \textsf{tc} tool to simulate local-area network~(LAN, RTT: 0.1~ms, 1~Gbps), metropolitan-area network~(MAN, RTT: 6~ms, 100~Mbps) and wide-area network~(WAN, RTT: 80~ms, 40~Mbps).
We set the computational security parameter $\symsec = 128$, which determines the key length of a pseudorandom function, and statistical security parameter $\statsec = 40$, with element size $k= 64$. 
Note we do not provide accuracy evaluation for \mostree as DTE only involves comparison, and there is no accuracy loss if the comparison is computed bit-wise in secure multiparty computation.





\subsection{Comparison with Three-party Works}
Since \mostree is the first three-party PDTE that considers honest majority security settings, we compare it with two latest three-party PDTE schemes~\cite{damgaard2019new,ji2021uc} with different security settings. 
The former~\cite{damgaard2019new} works in a dishonest majority setting whose security is a bit stronger than \mostree and the latter~\cite{ji2021uc} works in a semi-honest setting. 
We re-run protocols in ~\cite{damgaard2019new} and ~\cite{ji2021uc}, which we name as \spdz and \ucdt, respectively. 
We can build two kinds of \mostree based on different OS protocols.
However, the online phases of them are identical. 
To this, we give two lines (RSS-tree and \mostree) in Fig.~\ref{fig:exp:offline_communi} for offline communication while we only give one line (\mostree) for online communication in Fig.~\ref{fig:exp:online_communi}. 
We additionally give the concrete performance of two OS protocols in Appendix~\ref{appendix:exp_os}.
Since the DPF and RSS-based OS protocol outperforms the pure RSS-based OS, \mostree uses the DPF and RSS-based OS protocol if there is no explicit statement in the following. 

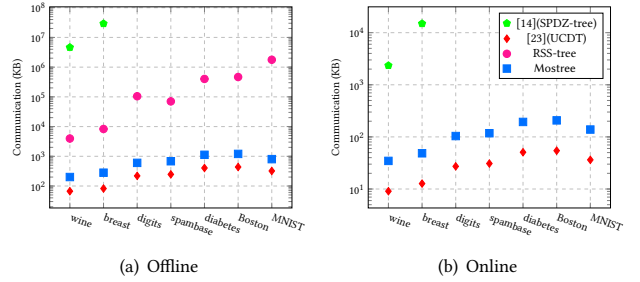
\begin{figure}[htbp]
    \centering
    \subfigure[Offline]{
        \begin{tikzpicture}[scale=0.47]
        \begin{axis}[%
        scatter/classes={%
        a={mark=oplus*,color=deeppink},b={mark=pentagon*,color=green},c={mark=diamond*,color=red},d={mark = square*,color=brandeisblue}
        },
        symbolic x coords={wine,breast,digits,spambase,diabetes,Boston,MNIST},
        x tick label style={anchor=north west,rotate=-20},
        ylabel=Number,
        scaled ticks=false, 
        legend style={font=\fontsize{5}{6}\selectfont},
        legend pos=outer north east, 
        grid=major,
        grid style=dashed,
        ymode=log,
        log basis y={10},
        ylabel={Communication (KB)},
        ]

        \addplot[mark=pentagon*,color=green,mark options={scale=1.5},scatter,only marks,scatter src=explicit symbolic]%
        coordinates {(wine,4630642.545)(breast,29175712.77)(digits,)(spambase,)(diabetes,)(Boston,)(MNIST,)};
        
        \addplot[mark=diamond*,color=red,mark options={scale=1.5},scatter,only marks,scatter src=explicit symbolic]%
        coordinates {(wine,66.8727)(breast,82.1376)(digits,219.238)(spambase,246.237)(diabetes,403.946)(Boston,435.563)(MNIST,321.164)};

        \addplot[mark=oplus*,color=deeppink,mark options={scale=1.5},scatter,only marks,scatter src=explicit symbolic]%
        coordinates {(wine,3964.91776)(breast,8298.73152)(digits,104748.2368)(spambase,70354.7392)(diabetes,400828.1088)(Boston,465650.2784)(MNIST,1772419.072)};

        \addplot[mark=square*,color=brandeisblue,mark options={scale=1.5},scatter,only marks,scatter src=explicit symbolic]%
        coordinates {(wine,200.5859375)(breast,280.8203125)(digits,601.7578125)(spambase,681.9921875)(diabetes,1123.28125)(Boston,1203.515625)(MNIST,802.34375)};


        \end{axis}
        \end{tikzpicture}
        \label{fig:exp:offline_communi}
    }
    \subfigure[Online]{
        \begin{tikzpicture}
        [scale=0.47]
        \begin{axis}[%
        scatter/classes={%
        a={mark=pentagon*,color=green},b={mark=diamond*,color=red},c={mark=oplus*,color=deeppink},d={mark = square*,color=brandeisblue}
        },
        symbolic x coords={wine,breast,digits,spambase,diabetes,Boston,MNIST},
        x tick label style={anchor=north west,rotate=-20},
        ylabel=Number,
        scaled ticks=false, 
        legend style={font=\fontsize{10}{11}\selectfont},
        legend pos=north east, 
        grid=major,
        grid style=dashed,
        ymode=log,
        log basis y={10},
        ylabel={Communication (KB)},
        ]
        \addplot[mark=pentagon*,color=green,mark options={scale=1.5},scatter,only marks,scatter src=explicit symbolic]%
        coordinates {(wine,2343.05536)(breast,14943.232)(digits,)(spambase,)(diabetes,)(Boston,)(MNIST,)};
        \addlegendentry{~\cite{damgaard2019new}(\spdz)}


        \addplot[mark=diamond*,color=red,mark options={scale=1.5},scatter,only marks,scatter src=explicit symbolic]%
        coordinates {(wine,9.07421875)(breast,12.70019531)(digits,27.19921875)(spambase,30.82421875)(diabetes,50.76171875)(Boston,54.38671875)(MNIST,36.26171875)};
        \addlegendentry{~\cite{ji2021uc}(\ucdt)}

        \addplot[mark=oplus*,color=deeppink,mark options={scale=1.5},scatter,only marks,scatter src=explicit symbolic]%
        coordinates {(wine,)(breast,)(digits,)(spambase,)(diabetes,)(Boston,)(MNIST,)};
        \addlegendentry{RSS-tree}

        \addplot[mark=square*,color=brandeisblue,mark options={scale=1.5},scatter,only marks,scatter src=explicit symbolic]%
        coordinates {(wine,34.63671875)(breast,48.47265625)(digits,103.8164063)(spambase,117.6523438)(diabetes,193.75)(Boston,207.5859375)(MNIST,138.40625)};
        \addlegendentry{\mostree}

        \end{axis}
        \end{tikzpicture} 
        \label{fig:exp:online_communi}
    }
\myfigspace
    \caption{Online and Offline Communication Cost. $y$-axis is in the logarithm scale. RSS-tree (\mostree) means we use pure RSS-based (DPF and RSS-based) OS protocol.}\label{fig:exp:communication}
\end{figure}

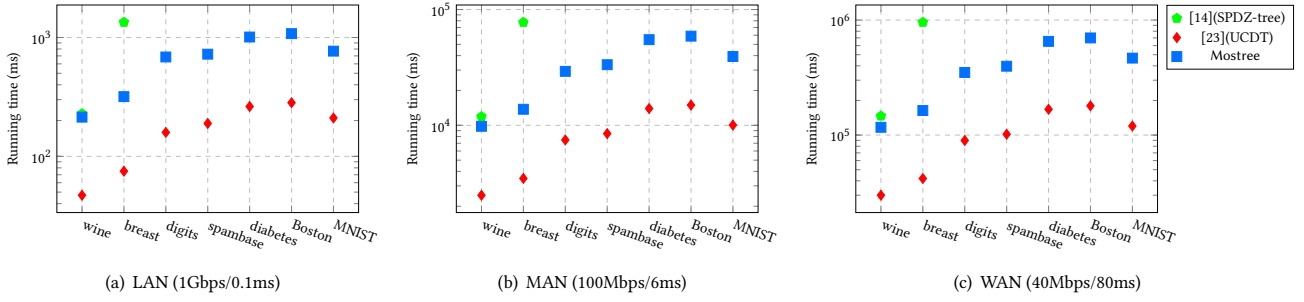
\begin{figure*}[htbp]
\centering
    \subfigure[LAN (1Gbps/0.1ms)]{
        \begin{tikzpicture}[scale=0.625]
        \begin{axis}[%
        scatter/classes={%
        a={mark=pentagon*,color=green},b={mark=diamond*,color=red},c={mark = square*,color=brandeisblue}
        },
        symbolic x coords={wine,breast,digits,spambase,diabetes,Boston,MNIST},
        x tick label style={anchor=north west,rotate=-20},
        ylabel=Number,
        scaled ticks=false, 
        legend pos= north west, 
        height=6cm,
        width=8cm,
        grid=major,
        grid style=dashed,
        ymode=log,
        log basis y={10},
        ylabel={Running time (ms)},
        ]
        
        \addplot[mark=pentagon*,color=green,mark options={scale=1.5},scatter,only marks,scatter src=explicit symbolic]%
        coordinates {(wine,228.398)(breast,1342.15)(digits,)(spambase,)(diabetes,)(Boston,)(MNIST,)};
        

        \addplot[mark=diamond*,color=red,mark options={scale=1.5},scatter,only marks,scatter src=explicit symbolic]%
        coordinates {(wine,47.042)(breast,74.9848)(digits,159.054)(spambase,189.125)(diabetes,262.627)(Boston,282.667)(MNIST,210.277)};


        \addplot[mark=square*,color=brandeisblue,mark options={scale=1.5},scatter,only marks,scatter src=explicit symbolic]%
        coordinates {(wine,213.732)(breast,317.93)(digits,685.479)(spambase,723.969)(diabetes,1009.12)(Boston,1077.49)(MNIST,768.051)};

        \end{axis}
        \end{tikzpicture}
    }
    \subfigure[MAN (100Mbps/6ms)]{
        \begin{tikzpicture}[scale=0.625]
        \begin{axis}[%
        scatter/classes={%
        a={mark=oplus*,color=deeppink},b={mark=pentagon*,color=green},c={mark=diamond*,color=red},d={mark = square*,color=brandeisblue}
        },
        symbolic x coords={wine,breast,digits,spambase,diabetes,Boston,MNIST},
        x tick label style={anchor=north west,rotate=-20},
        ylabel=Number,
        scaled ticks=false, 
        legend pos=north west, 
        height=6cm,
        width=8cm,
        grid=major,
        grid style=dashed,
        ymode=log,
        log basis y={10},
        ylabel={Running time (ms)},
        ]
        

        \addplot[mark=pentagon*,color=green,mark options={scale=1.5},scatter,only marks,scatter src=explicit symbolic]%
        coordinates {(wine,11890.4)(breast,77592)(digits,)(spambase,)(diabetes,)(Boston,)(MNIST,)};
        

        \addplot[mark=diamond*,color=red,mark options={scale=1.5},scatter,only marks,scatter src=explicit symbolic]%
        coordinates {(wine,2488.02)(breast,3479.14)(digits,7477.08)(spambase,8478.8)(diabetes,13960.4)(Boston,14954.4)(MNIST,10044.6)};



        \addplot[mark=square*,color=brandeisblue,mark options={scale=1.5},scatter,only marks,scatter src=explicit symbolic]%
        coordinates {(wine,9804.8)(breast,13760.2)(digits,29260.1)(spambase,33375.3)(diabetes,54916.8)(Boston,58909.7)(MNIST,39303.1)};

        \end{axis}
        \end{tikzpicture}

    }
    \subfigure[WAN (40Mbps/80ms)]{
        \begin{tikzpicture}[scale=0.625]
        \begin{axis}[%
        scatter/classes={%
        a={mark=pentagon*,color=green},b={mark=diamond*,color=red},c={mark = square*,color=brandeisblue}
        },
        symbolic x coords={wine,breast,digits,spambase,diabetes,Boston,MNIST},
        x tick label style={anchor=north west,rotate=-20},
        ylabel=Number,
        scaled ticks=false, 
        legend pos=  outer north east, 
        height=6cm,
        width=8cm,
        grid=major,
        grid style=dashed,
        ymode=log,
        log basis y={10},
        ylabel={Running time (ms)},
        ]
        

        \addplot[mark=pentagon*,color=green,mark options={scale=1.5},scatter,only marks,scatter src=explicit symbolic]%
        coordinates {(wine,146517)(breast,956410)(digits,)(spambase,)(diabetes,)(Boston,)(MNIST,)};
        \addlegendentry{~\cite{damgaard2019new}(\spdz)}
        

        \addplot[mark=diamond*,color=red,mark options={scale=1.5},scatter,only marks,scatter src=explicit symbolic]%
        coordinates {(wine,29970)(breast,41898.6)(digits,89754.2)(spambase,101733)(diabetes,167550)(Boston,179574)(MNIST,119830)};
        \addlegendentry{~\cite{ji2021uc}(\ucdt)}


        
        \addplot[mark=square*,color=brandeisblue,mark options={scale=1.5},scatter,only marks,scatter src=explicit symbolic]%
        coordinates {(wine,116868)(breast,163543)(digits,350294)(spambase,397041)(diabetes,653177)(Boston,699424)(MNIST,467116)};
        \addlegendentry{\mostree}
        
        \end{axis}
        \end{tikzpicture}
    }

\myfigspace
    \caption{Online Runtime in LAN/MAN/WAN Setting. $y$-axis is in the logarithm scale. 
    }\label{fig:exp:online_running_time}   
\end{figure*}

\noindent \textbf{Communication Evaluation.}
Fig.~\ref{fig:exp:communication} reports the communication of \mostree, \spdz and \ucdt. We obtain the communication for all listed datasets under different protocols except \spdz as we fail to compile the protocol over large trees. Its linear complexity makes it run out of memory during execution.

As we can see, over online communication, \mostree significantly outperforms~\spdz. Specifically, \spdz requires the most communication cost because it performs oblivious attribute selection and node comparison for each inner tree node. 
It is worth noticing that \mostree only shows an acceptable increase in online and offline communication overhead compared with \ucdt (\eg, \mostree requires ${\thicksim}4\times$ online communication for MNIST). In the offline phase, it is clear that RSS-tree (employs pure RSS-based OS) requires more communication resources than \mostree.

\begin{table*}[t!]
 \small
    \caption{Comparison with Two-party Protocols over Different Datasets and Networks} 
	\label{tab:exp:comparison with two-party}
    \vspace{-2mm}
	\centering
    \begin{threeparttable}
    \begin{tabular}{ccccccccccc}
    \toprule
          \multirow{2}{*}{\textbf{Protocols}}  &   \multirow{2}{*}{\textbf{Security}} & \multicolumn{3}{c}{\textbf{wine}}   & \multicolumn{3}{c}{\textbf{digits}} & \multicolumn{3}{c}{\textbf{MNIST}} \\ 
          \cmidrule(lr){3-5}\cmidrule(lr){6-8}\cmidrule(lr){9-11}
         & & LAN.T & WAN.T& Comm. & LAN.T & WAN.T& Comm. & LAN.T & WAN.T & Comm. \\
         \midrule
          Kiss \etal \cite{kiss2019sok}[GGG] & \Circle & \textbf{10.1} & \textbf{257.6} & 44.5 & 144.9 & \textbf{820.7} & 1499.2 & 2320.1 & 7088.4 &  23431.4\\
          Kiss \etal \cite{kiss2019sok}[HHH] & \Circle & 263.7 & 2083.5 & 90.2 & 1306.8 & 23554.6 & 990.7 & 13570.2 & 311726.9 & -\\
          Ma \etal \cite{ma2021let} & \Circle & 10.3 & 1300.4 & \textbf{1} & \textbf{28.2} & 4211.3 & 110.8 & \textbf{35.4} & \textbf{6493.5} & 270.3 \\
          Bai \etal \cite{bai2022scalable} & \Circle & 54.8 & 6295.8 & 336.6 & 152.1 & 18755.8 & 109.4 & 203.9 & 24997.3 & 369.4 \\
          Mostree  & \CIRCLE & 213.7 & 116868 & 34.6 & 685.5 & 350294 & \textbf{103.8} & 768.1 & 467116 & \textbf{138.4} \\         
        \bottomrule
    \end{tabular}
    \itshape \rm { {LAN.T}, {WAN.T}, and {Commu.} represent online running time under LAN, online running time under WAN, and online communication, respectively. \Circle: semi-honest, \CIRCLE: malicious. All running times are reported in milliseconds (ms), and communication in KB. The minimum value in each column is bolded.}
 \end{threeparttable}

\end{table*}

\begin{table}
\footnotesize
    \caption{Total Runtime (s) of \mostree under Different Network Conditions}\label{tab:exp:totalrunningtime}
    \vspace{-2mm}
    \centering
    \begin{tabular}{c|c|c|c|c|c|c|c}
    \hline
            & wine    & breast  & digits   & spambase & diabetes & Boston   & MNIST    \\ \hline
        LAN & 0.37 & 0.56 & 1.21 & 1.27 & 1.74 & 1.87 & 1.29 \\
        MAN & 16.90 & 23.70 & 50.81 & 57.79 & 94.87 & 101.61 & 67.95 \\
        WAN & 202.65  &  283.56 & 607.77  &  688.81  &  1133.29 &  1213.21  &  934.52\\ \hline
\end{tabular}
\end{table}


\begin{table}
\footnotesize
    \caption{\mostree Running on Trees with Varies Depth}
	
	\label{tab:exp:scalability}
    \vspace{-2mm}
	\centering
    \begin{tabular}{c|c|c|c|c|c|c|c}
    \hline
                Tree Depth & 20  & 25  & 30   & 35 & 40 & 45   & 50    \\ \hline
        TRT (ms) & 1397.9 & 1745.9 & 2046.3 & 2537.7 & 2923.1 & 2829.4 & 3270.6 \\
        TC (KB) & 362.0 & 446.3 & 530.6 & 614.9 & 699.1 & 783.4 & 594.3 \\ \hline
\end{tabular}
\itshape \rm {TRT and TC represent total running time and total communication, respectively.}
\end{table}

\noindent \textbf{Running Time.}
We evaluate \mostree, \spdz and \ucdt under LAN, MAN, and WAN network settings in Fig.~\ref{fig:exp:online_running_time}. 
Mostree shows a lower running time in the LAN setting than \spdz since \mostree only requires to perform oblivious node selection once for each tree level. The significant difference between \mostree and \spdz continues in the MAN and WAN network settings.
Compared with \ucdt, \mostree reports a slightly more running time (\eg, ${\thicksim}4\times$) for all listed trees in all network settings. 


Table~\ref{tab:exp:totalrunningtime} illustrates the total evaluation time of \mostree for different trees. Although the total running time of \mostree is linear to the tree size, the results in Table~\ref{tab:exp:totalrunningtime} shows it is more related to the tree depth. We acknowledge the running time is very sensitive to the latency. It can be highly improved by using batching techniques to send independent data together. We leave this as our future work.

\noindent \textbf{Scalability.}
We evaluate the scalability of \mostree in a LAN setting.
Following the approach in~\cite{tueno2019private}, we set the tree node number $m = 25d$ and vary $d$ from 20 to 50. Table~\ref{tab:exp:scalability} shows the evaluation result. 
We can see that while the tree depth increases, the communication grows slowly and linearly to the tree size, showing the scalability of \mostree. As for the running time, since the tree node number is linear to the tree depth in our setting, it also rises with the depth increase.

\subsection{Comparison with Two-party Works}
Most of the existing PDTE protocols focus on two-party settings. In particular, 
we compare \mostree with a comprehensive work~\cite{kiss2019sok} and two latest efficient protocols~\cite{ma2021let,bai2022scalable}. In particular, both communication and computation in work~\cite{kiss2019sok} are linear to the tree size, while the communication in~\cite{ma2021let,bai2022scalable} is sublinear.
Table~\ref{tab:exp:comparison with two-party} reports the online results. 
Since work~\cite{kiss2019sok} constructs PDTE modularly either using GC or HE, we compare \mostree with its two protocols: computation-efficient GGG and communication-efficient HHH.
The table shows that GGG in \cite{kiss2019sok} enjoys the best running time, and the work in \cite{ma2021let} shows the best communication for small trees like wine. 
While it goes to big trees like MNIST, \mostree outruns others in communication, including HHH. In particular, \mostree saves ${\thicksim}3\times$ communication cost than the latest sublinear work~\cite{bai2022scalable}. 
The main reason is that work~\cite{bai2022scalable} requires a two-party secure PRF evaluation for each node selection, which incurs significant communication overhead. 
In contrast, \mostree performs node selection mainly from cheap local evaluation, thus achieving better performance. 
We do acknowledge that this is partially due to the three-party honest majority setting, where \mostree exploits to design a more efficient PDTE. 

As for the running time, \mostree shows worse in some cases, but please note that \mostree achieves a much stronger security guarantee than other protocols. 
In some cases, GGG in \cite{kiss2019sok} shows the best running time. This is because it is evaluated purely based on the Garbled Circuit with the high communication price. As the tree size grows, work \cite{ma2021let} outperforms others, yet it leaks access patterns. 
In Table \ref{tab:exp:comparison with two-party}, we only compare \mostree with semi-honest two-party because the existing two-party work \cite{tai2017privacy} \cite{wu2016privately} only supporting malicious FO can be seen as the malicious version of HHH in \cite{kiss2019sok} (with additional ZKP and COT), with worse running time. 
We acknowledge that high computation in \mostree also means high monetary costs for cloud-assisted applications. 
Reducing the computation overhead further is our future work.



 \myspace
\section{Conclusion} \label{sec:conclusion}
This paper presents \mostree, a communication-efficient PDTE protocol. 
\mostree proposes two OS protocols with low communication by using RSS sharings and distributed point function. 
\mostree carefully combines oblivious selection protocols with data structure and lightweight consistency check techniques, achieving malicious security and sublinear communication. 
Our experiment results show that \mostree is efficient and practical. 

\myindent\textbf{Limitations and Future Works}. 
\mostree achieves sublinear online communication but still requires (super) linear computation cost. Achieving low computation overhead is also important, which we leave as our future work. 
In addition, we will extend malicious OS to other privacy-preserving applications, \eg, privacy-preserving inference using graph-based models. 


\section*{Acknowledgement} \label{sec:acknowledgement}
We thank the anonymous reviewers for their insightful comments and suggestions. 
Bai and Russello would like to acknowledge the MBIE-funded programme STRATUS (UOWX1503) for its support and inspiration for this research.
This research is supported by the National Research Foundation, Singapore under its Strategic Capability Research Centres Funding Initiative. Any opinions, findings and conclusions or recommendations expressed in this material are those of the author(s) and do not reflect the views of National Research Foundation, Singapore.

\bibliographystyle{ACM-Reference-Format}
\bibliography{ref}

\appendix

{
\section{Security Definition} \label{appendix:security-def}

\begin{definition}[DPF Security]
\label{Def:dpfsec}
	A two-party DPF satisfies the following requirements:
	\begin{itemize}[leftmargin=10pt]
		\item $\textbf{\rm Correctness}$: for any point function $f: \mathcal{D} \rightarrow \mathcal{R}$ and every $x \in \mathcal{D}$, if $(\fsskey{0}, \fsskey{1}) \leftarrow \textsf{\rm Gen} (1^{\symsec}, f)$ then ${\rm Pr} [\textsf{\rm Eval}(\fsskey{0}, x)+\textsf{\rm Eval}(\fsskey{1}, x)=f(x)] = 1$.
		
		\item $\textbf{\rm Secrecy}$: For any two point functions $f, f^*$, it holds that
		$\{ \fsskey{b}:(\fsskey{0}, \fsskey{1}) \leftarrow \textsf{\rm Gen}(1^{\symsec}, f) \} \stackrel{c}{\equiv} \{ \fsskeyy{b} : (\fsskeyy{0}, \fsskeyy{1}) \leftarrow \textsf{\rm Gen}(1^{\symsec}, f^*) \}$ for $b \in \{0,1\}$. 
		
	\end{itemize}
\end{definition}


\myspace 
\begin{definition}[Three-party secure computation] \label{def:malicous security}
    Let $\mathcal{F}$ be a three-party functionality. 
    A protocol $\Pi$ securely computes $\mathcal{F}$ with abort in the presence of one malicious party, if for every party $P_i$ corrupted by a \textit{probabilistic polynomial time} (PPT) adversary $\ADV$ in the real world, there exists a PPT simulator $\SIM$ in the ideal world with $\mathcal{F}$, such that,
\[\{\mathsf{IDEAL}_{\mathcal{F}, \SIM(z),i}(x_0, x_1, x_2, n)\} 
\overset{c}{\equiv} 
\{ \mathsf{REAL}_{\Pi, \ADV(z),i}(x_0, x_1, x_2, n) \}.\]
    where $x_i \in \{0,1\}^*$ is the input provided by $\party_i$ for $i \in [3]$, and $z \in \{0,1\}^*$ is the auxiliary information that includes the public input length information $\{|x_i|\}_{j \in [3]}$. 
    The protocol $\Pi$ securely computes $\mathcal{F}$ with abort in the presence of one malicious party with statistical error $2^{-\statsec}$ if there exists a negligible function $\mu(\cdot)$ such that the distinguishing probability of the adversary is less than $2^{-\statsec}+\mu(\symsec)$. 
\end{definition}

\section{Subprotocols used in \mostree}\label{appendix:ideal-fun}
This section shows protocols used in \mostree, including protocols that securely compute assumed RSS-based functionalities in section~\ref{sec:background} and others used in \mostree, \eg, secure RSS-based equality comparison.

\myindent\textbf{Protocol \protocol{rand} for generating a random RSS sharing}. 
Fig.~\ref{Protocol:rand} shows protocol $\protocol{rand}$ for generating an RSS sharing $\HMboolshare{r}$ for a random value $r \in \FF$. The protocol can be done non-interactively using PRF $F$ after a one-time setup for sharing PRF keys.

\begin{figure}
    \framebox{\begin{minipage}{0.93\linewidth} 
        
                {\bf Parameters}: Three parties $\{\party_0, \party_1, \party_2\}$; a field $\FF$ over which the RSS sharing works; a PRF $F: \mathcal{K} \times \mathcal{D} \rightarrow \mathcal{R}$ where $\mathcal{K} = \mathcal{D} = \{0,1\}^{\symsec}$ and $\mathcal{R} = \FF$. 

                {\bf [Setup]} Upon input $(\mathsf{setup})$, do:	
                \begin{enumerate} [leftmargin=10pt]
                    \item For $i \in [3]$, $\party_i$ samples a PRF key $\prfkey{i} \in \mathcal{K}$ and sends $\prfkey{i}$ to $\party_{i-1}$. $\party_{i}$ also receives $\prfkey{i+1}$ from $\party_{i+1}$.  
                \end{enumerate}
                {\bf [Rand]} Upon input $(\mathbf{rand}, \mathsf{sid})$, do:	
                
                \begin{enumerate}[leftmargin=10pt]                  
                    \item for $i \in [3]$, $\party_i$ defines $\sh{r}{i} \leftarrow F(\prfkey{i}, \mathsf{sid})$ and $\sh{r}{i+1} \leftarrow F(\prfkey{i+1}, \mathsf{sid})$. 
                    $\party_i$ defines $\HMboolshare{r}_i \leftarrow (\sh{r}{i}, \sh{r}{i+1})$. 

                    \item The parties output $\HMboolshare{r}$.                    
                    
                \end{enumerate}  
            \end{minipage} 
	} 
	\centering \caption{Protocol $\protocol{rand}$ for Securely Compute $\f{rand}$}\label{Protocol:rand}
\end{figure}

\myindent\textbf{Protocol $\protocol{open}$ for securely opening an RSS sharing}.
We show $\protocol{open}$ in Fig.~\ref{Protocol:open} for securely opening an RSS sharing. 
The intuition is that each party $\party_i$ will receive the share $\sh{x}{i+1}$ from both $\party_{i-1}$ and $\party_{i+1}$, which allows $\party_{i}$ to do cross-check and abort if the values are unequal. 

\begin{figure}
    \framebox{\begin{minipage}{0.93\linewidth} 
                {\bf Parameters}: Three parties $\{\party_0, \party_1, \party_2\}$; a field $\FF$ over which the RSS sharing works; a PRF $F: \mathcal{K} \times \mathcal{D} \rightarrow \mathcal{R}$ where $\mathcal{K} = \mathcal{D} = \{0,1\}^{\symsec}$ and $\mathcal{R} = \FF$. 
                
        	{\bf [Open]} Upon input $(\mathsf{sid}, \HMboolshare{x})$, do:	
                \begin{enumerate}[leftmargin=10pt] 
                    \item For $i \in [3]$, $\party_{i}$ sends $\sh{x}{i}$ to $\party_{i-1}$ and $\sh{x}{i-1}$ $\party_{i+1}$.

                    \item For $i \in [3]$, $\party_{i}$ receives $\sh{x}{i+1}$ from $\party_{i+1}$ and $\party_{i+2}$. If $\party_{i+1}$ and $\party_{i+2}$ send different values of $\sh{x}{i+1}$, send abort to all other parties. 
                    
                    \item If the protocol does not abort, for $i\in [3]$, $\party_i$ computes $x = \sum_{i \in [3]} \sh{x}{i}$ and outputs $x$.
                \end{enumerate}  
            \end{minipage} 
	} 
	\centering \caption{Protocol $\protocol{open}$ for Securely Compute $\f{open}$}\label{Protocol:open}
\end{figure}

\myindent\textbf{Protocol $\protocol{coin}$ for securely generating random coins}. 
Protocol $\protocol{coin}$ in Fig.~\ref{Protocol:coin} outputs a random value $r \in \FF$ to all the parties.
The idea is that the parties first call $\f{rand}$ to generate a random $\HMboolshare{r}$ and then open $\HMboolshare{r}$ using $\f{open}$.  

\begin{figure}
    \framebox{\begin{minipage}{0.93\linewidth} 
                {\bf Parameters}: Three parties $\{\party_0, \party_1, \party_2\}$; a field $\FF$ over which the RSS sharing works; a PRF $F: \mathcal{K} \times \mathcal{D} \rightarrow \mathcal{R}$ where $\mathcal{K} = \mathcal{D} = \{0,1\}^{\symsec}$ and $\mathcal{R} = \FF$. 
                
        	{\bf [Coin]} Upon input $(\mathsf{sid})$, do:	
                \begin{enumerate}[leftmargin=10pt] 
                    \item The parties call $\f{rand}$ to generate a random RSS sharing $\HMboolshare{r}$. 

                    \item The parties call $\f{open}$ to open $\HMboolshare{r}$. If the open does not abort, each party outputs $r$. 
                    
                \end{enumerate}  
            \end{minipage} 
	} 
	\centering \caption{Protocol $\protocol{coin}$ for Securely Compute $\f{coin}$}\label{Protocol:coin}
\end{figure}

\myindent\textbf{Protocol $\protocol{recon}$ for secure secret reconstruction}. 
Protocol $\protocol{recon}$ in Fig.~\ref{Protocol:recon} reconstructs a consistent RSS sharing $\HMboolshare{x}$ to $\party_i$. 
To check whether the reconstruction is correct or not, $\party_i$ receives $\HMboolshare{x}_{i+1}$ from both $\party_{i+1}$ and $\party_{i-1}$, and abort if the two shares are unequal.  

\begin{figure}
    \framebox{\begin{minipage}{0.93\linewidth} 
                {\bf Parameters}: Three parties $\{\party_0, \party_1, \party_2\}$; a field $\FF$ over which the RSS sharing works; a PRF $F: \mathcal{K} \times \mathcal{D} \rightarrow \mathcal{R}$ where $\mathcal{K} = \mathcal{D} = \{0,1\}^{\symsec}$ and $\mathcal{R} = \FF$. 
                
        	{\bf [Recon]} Upon input $(\mathsf{sid}, \HMboolshare{x}, i)$, do:	
                \begin{enumerate} [leftmargin=10pt]
                    \item $\party_{i}$ receives $\sh{x}{i+1}$ from $\party_{i+1}$ and $\party_{i-1}$. If $\sh{x}{i+1}$ from $\party_{i+1}$ and $\party_{i-1}$ do not match, abort. 

                    \item Compute $x = \sum_{j\in [3]} \sh{x}{j}$.  
                    
                \end{enumerate}  
            \end{minipage} 
	} 
	\centering \caption{Protocol $\protocol{recon}$ for Securely Compute $\f{recon}$}\label{Protocol:recon}
\end{figure} 

\myindent\textbf{Protocol $\protocol{share}$ for sharing a secret from $\party_i$}. 
Fig.~\ref{Protocol:share} shows the protocol $\protocol{share}$. It shares a secret $x$ from $\party_i$ among three parties.
In particular, $\protocol{share}$ first invokes $\f{rand}$ to generate a random RSS sharing $\HMboolshare{r}$ between the parties. Then the parties securely reconstruct $r$ to the party $\party_i$.
$\party_i$ broadcasts $\delta = x - r$ to the other two parties.
The other two parties cross-check whether they receive the same $\delta$. If yes, the parties jointly compute $\HMboolshare{x} \leftarrow \HMboolshare{r} +\delta$. 

\begin{figure}
    \framebox{\begin{minipage}{0.93\linewidth} 
                {\bf Parameters}: Three parties $\{\party_0, \party_1, \party_2\}$; a field $\FF$ over which the RSS sharing works; a PRF $F: \mathcal{K} \times \mathcal{D} \rightarrow \mathcal{R}$ where $\mathcal{K} = \mathcal{D} = \{0,1\}^{\symsec}$ and $\mathcal{R} = \FF$. 
                
        	{\bf [Share]} Upon input $(\mathsf{sid}, {x}, i)$, do:	
                \begin{enumerate} [leftmargin=10pt]
                    \item The parties call $\f{rand}$ to generate a random RSS sharing $\HMboolshare{r}$. 

                    \item The parties call $\f{recon}$ over $\HMboolshare{r}$ and reconstruct $r$ to $\party_{i}$. $\party_i$ broadcasts $\delta = x - r$ to the other parties.    

                    \item $\party_{i+1}$ and $\party_{i+2}$ check if they receive the same $\delta$. If not, abort.  

                    \item The parties output $\HMboolshare{x} \leftarrow \HMboolshare{r} + \delta$. 
                \end{enumerate}  
            \end{minipage} 
	} 
	\centering \caption{Protocol $\protocol{share}$ for Securely Compute $\f{share}$}\label{Protocol:share} 
\end{figure} 

\myindent\textbf{Protocol \protocol{mut} for multiplication}. 
Fig.~\ref{Protocol:mut} shows how to perform multiplication with up to an additive attack.  
This protocol is essentially a semi-honest multiplication protocol. 
We note a malicious party can add an error $e$ in the resharing phase to produce an incorrect RSS sharing $\HMboolshare{x\cdot y +e}$.

\begin{figure}
    \framebox{\begin{minipage}{0.93\linewidth} 
                {\bf Parameters}: Three parties $\{\party_0, \party_1, \party_2\}$; a field $\FF$ over which the RSS sharing works; a PRF $F: \mathcal{K} \times \mathcal{D} \rightarrow \mathcal{R}$ where $\mathcal{K} = \mathcal{D} = \{0,1\}^{\symsec}$ and $\mathcal{R} = \FF$. 
                
        	{\bf [Mul]} Upon input $(\mathsf{sid}, 
        \HMboolshare{x}, \HMboolshare{y})$, do:	
                \begin{enumerate} [leftmargin=10pt]
                    \item For $i \in [3]$, $\party_i$ computes $\sh{t}{i} \leftarrow \sh{x}{i}\cdot \sh{y}{i} + \sh{x}{i-1} \cdot \sh{y}{i} + \sh{x}{i}\cdot \sh{y}{i-1}$. 
                    $(\sh{t}{0}, \sh{t}{1}, \sh{t}{2})$ forms a $(^3_3)$-sharing $\Twoboolshare{t}$.

                    \item The parties call $\f{rand}$ to generate a random RSS sharing $\HMboolshare{r}$. 

                    \item For $i \in [3]$, $\party_i$ computes $\sh{z}{i} \leftarrow \sh{t}{i} + \sh{r}{i} - \sh{r}{i-1}$. $\party_{i}$ sends $\sh{z}{i}$ to $\party_{i+1}$ and receives $\sh{z}{i-1}$ from $\party_{i-1}$. 

                    \item For $i \in [3]$, $\party_{i}$ defines $\HMboolshare{z}_i = (\sh{z}{i}, \sh{z}{i-1})$.  
                \end{enumerate}  
            \end{minipage} 
	} 
	\centering \caption{Protocol $\protocol{mut}$ for Securely Compute $\f{mut}^{\FF}$}\label{Protocol:mut} 
\end{figure} 

\myindent\textbf{Protocol \protocol{CheckZero} }.
In Fig.~\ref{Protocol:checkzero}, we show a protocol \protocol{CheckZero}. It can check whether an RSS sharing $\HMboolshare{x}$ is a sharing of 0.
In particular, the protocol first calls $\f{rand}$ to generate a random RSS sharing $\HMboolshare{r}$. The parties call $\f{mut}^{\FF}$ to compute $\HMboolshare{w} \leftarrow \HMboolshare{x}\cdot \HMboolshare{r}$. 
The parties then open $\HMboolshare{w}$ via $\f{open}$ and check if $w = 0$ and abort if not the case. 

\begin{figure}
    \framebox{\begin{minipage}{0.93\linewidth} 
                {\bf Parameters}: Three parties $\{\party_0, \party_1, \party_2\}$; a field $\FF$ over which the RSS sharing works; a PRF $F: \mathcal{K} \times \mathcal{D} \rightarrow \mathcal{R}$ where $\mathcal{K} = \mathcal{D} = \{0,1\}^{\symsec}$ and $\mathcal{R} = \FF$. 
                
        	{\bf [CheckZero]} Upon input $(\mathsf{sid}, 
        \HMboolshare{x})$, do:	
                \begin{enumerate} [leftmargin=10pt]
                    \item The parties call $\f{rand}$ to generate a random RSS sharing $\HMboolshare{r}$. 

                    \item The parties compute $\HMboolshare{w} \leftarrow \HMboolshare{x} \cdot \HMboolshare{r}$ by calling $\f{mul}^{mul}$.  

                    \item The parties call $\f{open}$ to open $\HMboolshare{w}$. If the open aborts or $w \ne 0$, return $\mathsf{Flase}$; otherwise return $\mathsf{True}$. 
                \end{enumerate}  
            \end{minipage} 
	} 
	\centering \caption{Protocol $\protocol{CheckZero}$ for Securely Compute $\f{CheckZero}$}\label{Protocol:checkzero} 
\end{figure} 

\myindent\textbf{Protocol $\protocol{eq}$ for secure equality test}. 
In Fig.~\ref{protocol:equality-test}, we use an equality test protocol $\protocol{eq}$ for securely comparing whether two shared secrets are equal or not, and share the equality test result between the parties. 
To achieve malicious security, the parties need to check the correctness of multiplications.
This can be done by the triple-based multiplication verification trick, where the triples used for verification are generated by a cut-and-choose method from~\cite{furukawa2017high}. 

\begin{figure}
\mysize
	\framebox{\begin{minipage}{0.99\linewidth}
	\small
	{\bf Input}:
	An RSS-sharing index $\HMboolshare{\idx}$ for $\idx \in [n]$ and a public value $j$. 
    {\bf Output}:
    An RSS-sharing $\HMboolshare{u[j]}$ where $u[j]$ is 1 if $idx == j$, otherwise $u[j]$ is 0.

    \begin{enumerate} [leftmargin=10pt]
        \item The parties naturally share the public value $j$ as $\HMboolshare{j} = (0,0,j)$ and parse $\HMboolshare{\idx} = \HMboolshare{\idx[l-1]}\cdots \HMboolshare{\idx[0]}$ and $\HMboolshare{j} = \HMboolshare{j[l-1]}\cdots \HMboolshare{j[0]}$.
    
        \item The parties perform $\HMboolshare{h[q]} \leftarrow \HMboolshare{\idx[q]} \oplus \HMboolshare{j[q]}$ for $q \in [l]$.
    
        \item Set $\HMboolshare{u[j]} \leftarrow \HMboolshare{h[0]}$.
    
        \item For $q \in \{1, \cdots, l-1\}$,
    
    \begin{enumerate} [leftmargin=10pt]
        \item The parties perform $\HMboolshare{u[j]} \leftarrow \HMboolshare{u[j]}\cdot \HMboolshare{h[q]}$.
    \end{enumerate}

    \end{enumerate}
		
	\end{minipage}}
    \myspace
	\caption{Equality Test $\protocol{eq}$}
	\label{protocol:equality-test}
\end{figure}

\section{Security Proof}\label{section:proof}
\myspace
\subsection{Proof of Theorem~\ref{thm:rss-os-security}}\label{proof:rss-sos}
\begin{proof}
For any PPT adversary $\ADV$, we construct a PPT simulator $\SIM$ that can simulate the adversary's view with accessing the functionalities $\f{rand}$, $\f{3pc}^{\FF_2}$ and $\f{open}$. In the cases where $\SIM$ aborts or terminates the simulation, $\SIM$ outputs whatever $\ADV$ outputs.

\myindent\textbf{Simulating preprocess phase}. $\SIM$ plays the role of $\f{rand}$ and uses the simulator of $\f{rand}$ for simultation. For each $j \in [m]$, $\SIM$ simulates $\f{3pc}^{\FF_2}$ in $\protocol{eq}$ and receives an error from the adversary. $\SIM$ aborts if $\ADV$ sends any non-zero error.

\myindent\textbf{Simulating selection phase}. We assume $\HMboolshare{{rdx}}$ and $\HMboolshare{\B{v}}$ are correctly shared/simulated in the first stage. $\SIM$ plays the role of $\f{open}$ and uses the simulator of $\f{open}$ for simulation and receives an error from the adversary. $\SIM$ aborts if $\ADV$ sends any non-zero error. For the local shifting and local multiplication computation, $\SIM$ is easy to simulate. Towards the re-sharing phase, the simulation randomly samples $\party_{i+2}$'s share and sends it to $\ADV$. In the real protocol execution, this is generated by a PRF $F$. Therefore, the simulation is computationally indistinguishable from real protocol execution due to the security of PRF. Overall, the simulation is indistinguishable from real-world execution.
\end{proof}

\subsection{Proof of Theorem~\ref{thm:sos-security}}\label{proof:sos}
\begin{proof}
For any PPT adversary $\ADV$, we construct a PPT simulator $\SIM$ that can simulate the adversary's view with accessing the functionality $\f{os}$. In the cases where $\SIM$ aborts or terminates the simulation, $\SIM$ outputs whatever $\ADV$ outputs. 

\myindent\textbf{Simulating preprocess phase}.
When the corrupted party $\party_i$ plays the role of DPF key generator, $\SIM$ plays the role of honest parties $\party_{i+1}$ and $\party_{i+2}$, and receives a pair of DPF keys from $\ADV$. $\SIM$ can check whether the keys are correct and abort if not. 
When an honest party takes the turn for DPF key generation, $\SIM$ generates a pair of DPF keys and samples a random share $r_i \rnd \ZZ_m$. $\SIM$ sends the key $(\fsskey{i}, r_i)$ to $\ADV$. $\SIM$ runs VDPF verification protocol with $\ADV$, and aborts if $\ADV$ aborts in the verification protocol. 
For all other messages sent from honest parties to the corrupted one, the simulator samples them randomly.

The above simulation is indistinguishable from real-world execution. First, if $\ADV$ sends incorrect DPF keys, then the real-world execution would have aborted due to VDPF check. 
In the ideal world, $\SIM$ can check keys directly, thus the ideal world aborts with indistinguishable probability.
When the corrupted party plays the role of DPF evaluator, the DPF keys are honestly generated by $\SIM$. If $\ADV$ follows the VDPF key verification protocol, the check will always pass, otherwise the check protocol will abort. Therefore, $\SIM$ just runs the check protocol with $\ADV$ as in the real-world protocol, and it can always abort with indistinguishable probability.

\myindent\textbf{Simulating selection phase}.
If the protocol does not abort after the preprocessing stage, then all DPF keys generated by $\ADV$ in $\SIM$ are correct. 
The only issue in the selection phase is whether $\ADV$ follows the protocol honestly. Also, $\SIM$ has to successfully extract the error term in the simulation.   
We assume $\HMboolshare{\tree}$ and $\HMboolshare{{\idx}}$ are already correctly shared/simulated in the first place, which means the honest parties hold correct and consistent RSS shares. 

For all local computations, $\SIM$ is easy to simulate. $\SIM$ only needs to simulate the view between honest parties and the corrupted party and to abort with indistinguishable probability. Note that only a few parts in the selection phase require interaction.  
The first part is on securely reconstructing $\Delta$. Since $\HMboolshare{\Delta} = \HMboolshare{\rdx} \oplus \HMboolshare{\idx}$ and both $\HMboolshare{\rdx}$ and $\HMboolshare{\idx}$ are correct RSS sharings, $\HMboolshare{\Delta}$ is also a consistent sharing by definition. 
This means that $\SIM$ can cross-check whether $\ADV$ sends the correct value using the share of honest party $\party_{i+2}$: 1) if $\ADV$ sends incorrect share, $\SIM$ just aborts; 2) if $\ADV$ sends the correct share, $\SIM$ samples a random $\Delta \in \ZZ_m$ and computes the shares of $\party_{i+1}$ and $\party_{i+2}$ according to $\Delta$ and the RSS shares of $\party_i$, and sends to $\ADV$. In this case, the parties open the random $\Delta$ to $\ADV$.  
Another part is from re-sharing the selection result from $\B{A}[\idx]$ from $(^3_3)$-sharing back to RSS sharing. Here $\ADV$ can add an error and $\SIM$ extracts the error as follows: $\SIM$ receives the share $z^*_i$ from the corrupted party to an honest party and updates $\party_{i+1}$'s local share correspondingly. $\SIM$ also locally computes value $z_i$, which is the correct value that the corrupted party should send~($\SIM$ can do this since he knows all necessary shares to compute $z_i$). Then, $\SIM$ computes $d \leftarrow z_i - z^*_i$ to $\f{os}$. $\SIM$ sends $d$ to $\f{os}$, completing simulation.   

Note that in simulating the open step of $\Delta$, $\SIM$ only uses shares of honest parties. These shares are either randomly simulated or computed from available local data. Also, they are consistent with the shares/data held by the corrupted party. Therefore, the simulation is perfectly indistinguishable from real protocol execution.
Towards re-sharing phase, the simulation randomly samples $\party_{i+2}$'s share and sends it to $\ADV$. In the real protocol execution, this is generated by a PRF $F$. Therefore, the simulation is computationally indistinguishable from real protocol execution due to the security of PRF. Overall, the simulation is indistinguishable from real-world execution. 
\end{proof}

\subsection{Proof of Theorem~\ref{thm:pdte-security}}\label{proof:pdte} 
\begin{proof}
For any PPT adversary $\ADV$ who corrupts party $\party_i$, we construct a PPT simulator $\SIM$ who can simulate the adversary's view with accessing the functionality $\f{os}$. In the cases where $\SIM$ aborts or terminates the simulation, $\SIM$ outputs whatever $\ADV$ outputs.

\myindent\textbf{Simulating setup}. 
$\SIM$ samples random shares for $\tree$ and hands the shares to $\ADV$. $\SIM$ also randomly samples shares for the honest parties in order to perform subsequent simulations.
$\SIM$ plays the role of $\f{rand}$ and uses the simulator of $\f{rand}$ for simulation. 
For $j \in [m]$, $\SIM$ plays the role of $\f{mul}^{\FF_{2^{\ell}}}$ and receives $d_j$ from the adversary. From $d_j$, $\SIM$ adds the error into the share of MAC value $\HMboolshare{ \sigma(\tree[j]) }$ for honest party $\party_{i+1}$. 
$\SIM$ simulates $\f{rand}$ and $\f{CheckZero}$ in $\protocol{MacCheck}$.
If there exists any $j \in [m]$ such that $d_j \ne 0$ or the simulation aborts from $\f{rand}$ and $\f{CheckZero}$ in $\protocol{MacCheck}$~($\SIM$ uses existing simulation strategy for $\f{rand}$ and $\f{CheckZero}$), $\SIM$ aborts. 

The above simulation is indistinguishable from real protocol execution. Since the protocol is designed in a hybrid model, existing simulation strategy for $\f{rand}$, $\f{mul}^{\FF_{2^{\ell}}}$, and $\f{CheckZero}$ are available, thus $\SIM$ uses them directly. 
$\SIM$ can also extract the strategy of $\ADV$ by receiving the error term in $\f{mul}^{\FF_{2^{\ell}}}$ thus $\SIM$ can abort correspondingly. Note that the MAC we use is over $\FF_{2^{\ell}}$. In real protocol execution, any additive error will incur MAC check fail except with probability $\frac{1}{{2^{\ell}} - 1}$. 
Overall, the above simulation is statistically indistinguishable from the real-world execution, with statistical error $\frac{1}{2^{\ell}-1}$.

\myindent\textbf{Simulating evaluation}.
For all bit-wise secure computation~(including inner product, comparison and MUX), $\SIM$ directly uses existing simulation strategy for $\f{3pc}^{\FF_2}$ directly. $\SIM$ plays the role of $\f{os}$ for each oblivious selection and receives an error term from $\ADV$. If $\SIM$ receives any non-zero error in simulating $\f{os}$, $\SIM$ aborts at the end of the protocol. 
The simulation for bit-wise computation using $\f{3pc}^{\FF_2}$ is well-studied thus $\SIM$ can directly use existing simulation strategy~(see \cite{furukawa2017high}). This part is indistinguishable from real-world execution.
The only issue is from $\f{os}$. Since $\SIM$ receives the error terms from $\ADV$ per selection, $\SIM$ just aborts if $\ADV$ sends any non-zero error. Overall, the above simulation is statistically indistinguishable from real-protocol execution, with statistical error $\frac{1}{2^{\ell}-1}$.
\end{proof}

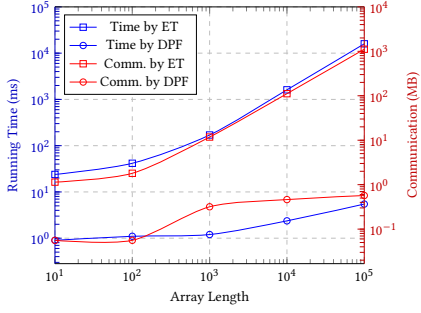
\begin{figure}[htbp]
    \centering
    \begin{tikzpicture}[scale=0.6]
    \pgfplotsset{
        legend pos=north west,
        y axis style/.style={
            yticklabel style=#1,
            ylabel style=#1,
            y axis line style=#1,
            ytick style=#1
       }
    }
    
    \begin{axis}[
      axis y line*=left,
      y axis style=blue!75!black,
      ymin= 0,
        ymax= 10^5,
        ymode=log,
        log basis y={10},
        xmin= 10,
        xmax= 10^5,
        xmode=log,
        log basis x={10},
      xlabel=Array Length,
      ylabel=Running Time (ms),
      grid=major,
        grid style=dashed,
    ]
    \addplot[smooth,mark=square,blue] 
      coordinates{
        (10,23.876)
        (100,41.652) 
        (1000,170.167)
        (10000,1611.28)
        (100000,15915.2) 
    };\label{first}
    
    \addplot[smooth,mark=o,blue] 
      coordinates{
        (10,0.9)
        (100,1.1) 
        (1000,1.2)
        (10000,2.37)
        (100000,5.47) 
    };\label{second}

    
    \end{axis}
    
    \begin{axis}[
      axis y line*=right,
      axis x line=none,
      ylabel= Communication (MB),
      y axis style=red!75!black,
      xmin= 10,
        xmax= 10^5,
        xmode=log,
        log basis x={10},
      ymin= 0,
        ymax= 10^4,
        ymode=log,
        log basis y={10},
    ]
    \addlegendimage{/pgfplots/refstyle=first}\addlegendentry{Time by ET}
    \addlegendimage{/pgfplots/refstyle=second}\addlegendentry{Time by DPF}

    \addplot[smooth,mark=square,red] 
      coordinates{
        (10, 1.14101)
        (100,1.80383) 
        (1000,11.7952)
        (10000,111.548)
        (100000,1115.96) 
    };
    \addlegendentry{Comm. by ET}
    
    \addplot[smooth,mark=o,red] 
      coordinates{
        (10,0.056)
        (100,0.056) 
        (1000,0.318)
        (10000,0.462)
        (100000,0.570) 

    };
    \addlegendentry{Comm. by DPF}
    
    \end{axis}
    
    \end{tikzpicture}
    \caption{Unit Vector Generation.  $y$-axis is in the logarithm scale. ET and Comm. represent equality test and communication cost, respectively.}
    \label{fig:unitvector}
\end{figure}

\subsection{Evaluation of Two OS Protocols}\label{appendix:exp_os}
We show the performance of unit vector generation of two proposed OS protocols in Fig.~\ref{fig:unitvector}. The RSS-based OS employs equality test to generate the unit vector, which requires three parties to jointly compare each index with a given value, resulting in linear computation and communication. 
The DPF-based OS needs to share the keys. Due to the sublinear property of the DPF scheme, the required communication is sublinear to the array length. However, to expand the keys to the unit vector, each party needs to perform local computation for each element, which also requires linear computation. Note both of these two unit vector generations can be moved offline.
}

\end{document}